\documentclass[preprint,amsmath,amssymb,aps,prb,floatfix, bibliography]{revtex4-1}
\usepackage{natbib}
\usepackage{units}
\usepackage{tabto}
\usepackage{color}
\usepackage{textcomp}
\usepackage{natmove}
\usepackage{mciteplus}
\usepackage[export]{adjustbox}
\usepackage{amsbsy}
\usepackage{amssymb}
\usepackage{subfloat}
\usepackage{float}
\usepackage{geometry}
\usepackage{setspace}
\usepackage{xkeyval}
\usepackage{graphicx}
\usepackage[caption=false]{subfig}
\usepackage{placeins}

\usepackage{tikz}
\usepackage{csquotes}
\usepackage{colortbl}
\usepackage{amsmath}
\usepackage{placeins}
\usepackage{graphicx}
\usepackage[unicode=true, bookmarks=true, bookmarksnumbered=false, bookmarksopen=false,
  breaklinks=false,pdfborder={0 0 0},backref=false,colorlinks=true]{hyperref}
 \hypersetup{pdftitle={Multiple Triple-Point Fermions in Heusler Compounds},
  pdfauthor={Ranjan Kumar Barik, Ravindra Shinde and Abhishek K. Singh},
  pdfsubject={Computational Material Science},colorlinks=true,
  unicode=false,pdftoolbar=true,pdfmenubar=true,pdffitwindow=true, pdfstartview={FitH}, pdfnewwindow=true, pdfcreator={Latex}, linkcolor=red, citecolor=blue, urlcolor=blue}

\DeclareGraphicsExtensions{.jpg, .eps, .png}

\definecolor{NLRed}{RGB}{215,24,30}
\definecolor{abstractcolor}{RGB}{255,243,201}

 \makeatother

\begin{document}

\title{Multiple Triple-Point Fermions in Heusler Compounds}

\author{Ranjan Kumar Barik}
\thanks{Contributed equally}
\author{Ravindra Shinde}
\thanks{Contributed equally}
\author{Abhishek K. Singh}
\email[Corresponding Author: ]{abhishek@iisc.ac.in}
\affiliation{Materials Research Center, Indian Institute of Science, Bangalore 560012, India}


\begin{abstract}
Using the density functional theoretical calculations, we report a new set of topological semimetals X$_{2}$YZ (X = \{Cu, Rh, Pd, Ag, Au, Hg\}, Y = \{Li, Na, Sc, Zn, Y, Zr, Hf, La, Pr, Pm, Sm, Tb, Dy, Ho, Tm\} and Z =\{Mg, Al, Zn, Ga, Y, Ag, Cd, In, Sn, Ta, Sm\}), which show the existence of multiple topological triple point fermions along four independent $C_{3}$ axes. These fermionic quasiparticles have no analogues elementary particle in the standard model. The angle-resolved photoemission spectroscopy is simulated to obtain the exotic topological surface states and the characteristic Fermi arcs. The inclusion of spin-orbit coupling splits the triple-point to two Dirac points. The triple-point fermions are exhibited on the easily cleavable (111) surface and are well separated from the surface $\bar{\Gamma}$ point, allowing them to be resolved in the surface spectroscopic techniques. This intermediate linearly dispersive degeneracy between Weyl and Dirac points may offer prospective candidates for quantum transport applications. 
\end{abstract}

\maketitle

\section{Introduction}

Quantum topological materials have been at the forefront of the intense research in condensed matter physics and materials science in recent years, as they exhibit fundamentally new physical phenomena with potential applications for novel devices \cite{Bernevig1757,MooreNature464.2010, PhysRevB.83.205101,PhysRevB.84.075129,PhysRevLett.107.186806, PhysRevB.87.245112}. The topological insulator, for instance, was the first three-dimensional topological material to be predicted and discovered\cite{MooreNature464.2010, hasan2010colloquium, RevModPhys.83.1057, PhysRevLett.98.106803, Hsieh919}. This was  subsequently followed by the experimental observations of Weyl semimetals (WSMs)\cite{PhysRevB.83.205101, BurBal11, PhysRevB.86.115208, weng2015weyl, xu2015discovery, Xu294, PhysRevLett.107.186806,HuaXuBel15}. The recent research in nontrivial topological materials has been focused on the gapless materials, because of linear band crossings and the existence of exotic fermions as quasiparticles. 

The topological semimetals (TSMs) are classified based on dimensionality and degeneracy of band crossings. A nodal point (zero-dimensional crossing) with two- and four-fold band degeneracy characterizes the Weyl and Dirac semimetals (DSMs), \cite{PhysRevLett.108.140405, wang2012dirac, PhysRevB.88.125427} respectively, while a nodal line (one-dimensional crossing) gives the corresponding nodal-line semimetals \cite{PhysRevLett.115.036806,yu2015topological}. The band crossing points of WSMs, known as Weyl nodes, have definite integer chirality, and they always appear in pairs. The WSMs also exhibit peculiar surface states, known as Fermi arcs, which connects a pair of surface-projected Weyl nodes with opposite chirality. DSMs with four-fold band degeneracy can be thought of as a special case of WSM with merged Weyl nodes with zero effective chirality. Nodal-line semimetals are considered as a precursor for other topological phases: they can evolve into Weyl points, convert into Dirac points, or become a topological insulator by the introduction of the spin-orbit coupling (SOC) or mass term \cite{Yu2016}. 

More recently, the classification mentioned above of fermions has been expanded to accommodate the triply-degenerate nodal points. This three-fold degeneracy gives rise to what is known as triple-point (TP) fermions\cite{PhysRevLett.116.186402,Bradlynaaf5037,PhysRevLett.117.076403,PhysRevB.93.241202,PhysRevX.6.031003,PhysRevB.94.165201,ChanXuHua17} and has been confirmed experimentally \cite{ChanXuHua17}. These peculiar TP fermions do not have analog in the standard model of the high energy physics, so the existing topological invariants will not be applicable in contrast to the Dirac and Weyl nodes.\cite{ChanXuHua17} Thus, these materials are expected to show new topological phenomena, transport behaviors, and spectroscopic responses, not present in DSMs and WSMs. 
The TP fermions having novel band crossings have triggered the search of quantum topological materials, mainly because of promising applications.
Many candidate materials in different space groups have been predicted to show this feature \cite{ChanXuHua17}, specifically with the crystal structures having three-fold rotational symmetry.  In this Communication, we predict a set of Heusler topological materials X$_2$YZ with characteristic TP fermions, which are present even in the absence of SOC. With the inclusion of SOC, the TP splits into two separate Dirac fermions. The two distinct observed TPs shift anti-parallel with the change in the row of constituent atoms offering tunability. 

\begin{figure*}[ht]
\subfloat[] {\label{subfig:crystal_and_BZ}\includegraphics[width=\textwidth, valign=b]{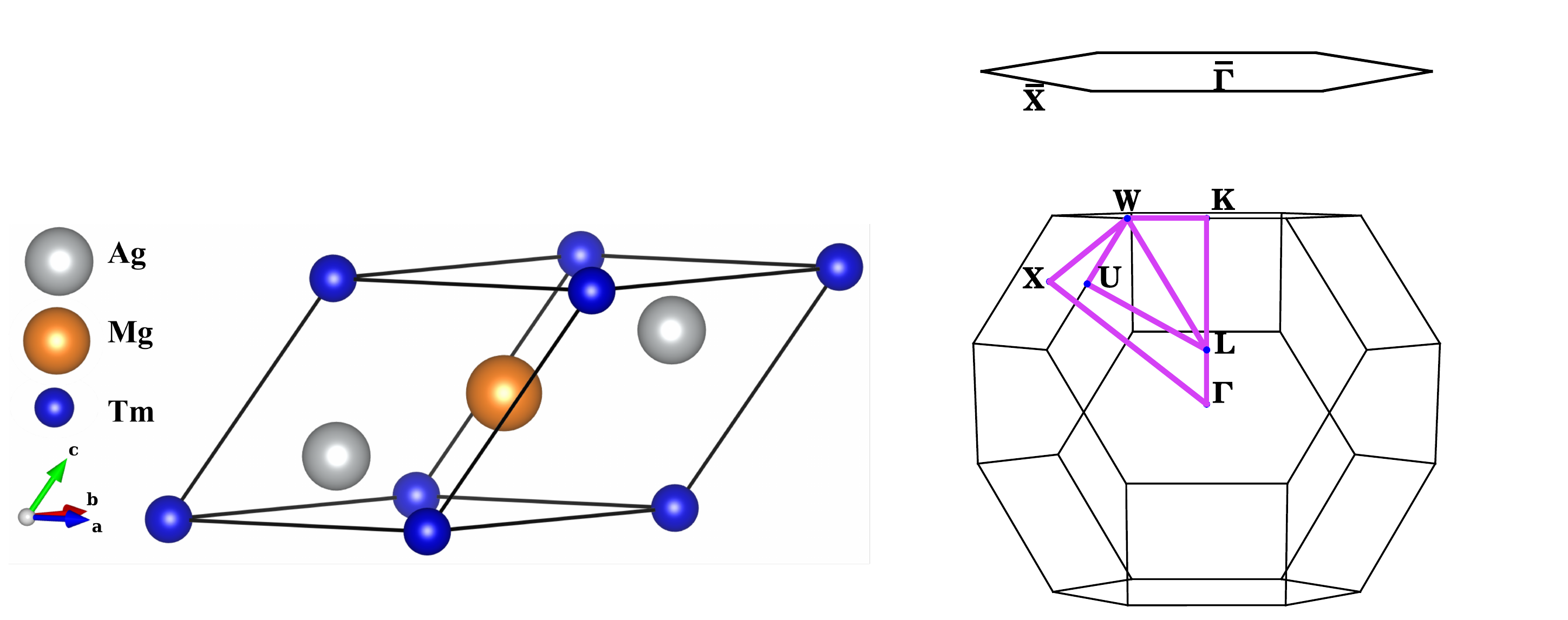}} \\
\subfloat[] {\label{subfig:bandstructure_nosoc}\includegraphics[width=0.5\textwidth]{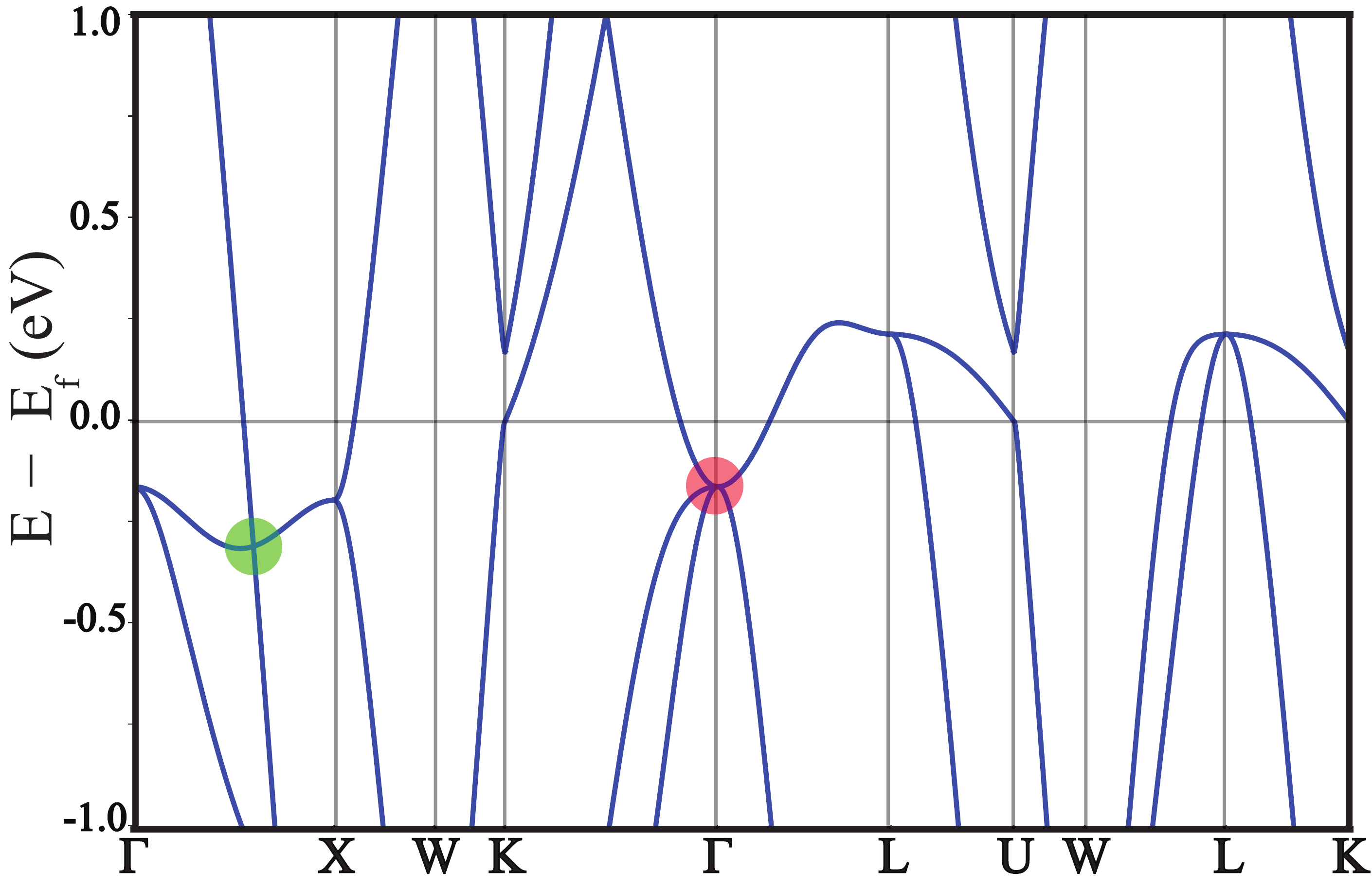}} 
\subfloat[] {\label{subfig:bandstructure_soc}\includegraphics[width=0.5\textwidth]{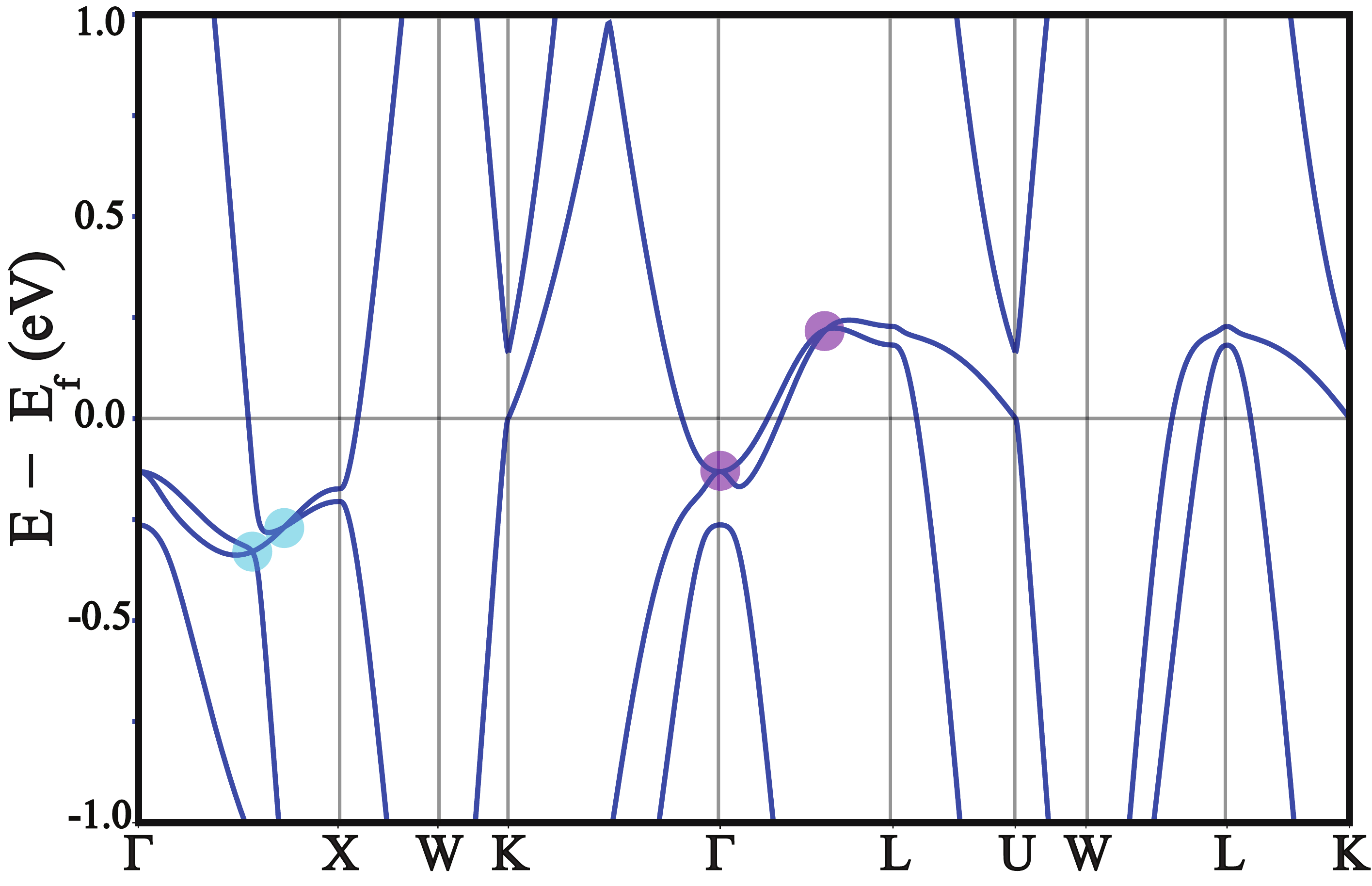}} 
\caption{(a) The crystal structure of Ag$_{2}$TmMg compound with space group $Fm\bar{3}m$ (No. 225). (b) The corresponding three-dimensional Brillouin zone along with its high-symmetry k-points and its projection onto the equivalent (111) surface. (c) The band structure of Ag$_{2}$TmMg without SOC and (d) with SOC.}
\label{fig:crystal_structure}
\end{figure*}

\section{\label{sec:theory}Methodology}

The calculations were performed using the first-principles density functional theory (DFT) \cite{kohn1965self} as implemented in the Vienna ab initio simulation package (VASP) \cite{kresse1996efficiency, kresse1996efficient}. Projector augmented wave (PAW) \cite{blochl1994projector,kresse1999ultrasoft} potentials were used to represent the ion-electron interactions. The electronic exchange and correlation were approximated by the generalized gradient approximation (GGA) using Perdew-Burke-Ernzerhof (PBE) type of functionals \cite{perdew1996generalized}. The wave functions were expanded in a plane wave basis with an energy cut-off of 400 eV and a $\Gamma$-centered 12$\times$12$\times$12 Monkhorst-Pack \cite{monkhorst1976special} \textbf{k}-grid Brillouin zone sampling. The calculations were done both with and without spin-orbit coupling. The Bloch spectral functions and the Fermi surfaces were calculated based on the iterative Green's function method \cite{sancho1985highly} by obtaining the tight-binding Hamiltonian from the maximally localized Wannier functions \cite{mostofi2014updated}, as implemented in the WannierTools package\cite{wann_tools}.

\section{\label{results-and-discussion}Results and Discussion}
\subsection{Crystal and electronic structure}

Ag$_{2}$TmMg belongs to the face-centered cubic centrosymmetric crystal structure (Fig. \protect\ref{fig:crystal_structure}\subref{subfig:crystal_and_BZ}) with $Fm\bar{3}m$ (No. 225) space group. Tm and Mg atoms occupy 8c ($\nicefrac{3}{4}$, $\nicefrac{3}{4}$, $\nicefrac{3}{4}$) and 8c ($\nicefrac{1}{4}$, $\nicefrac{1}{4}$, $\nicefrac{1}{4}$) Wyckoff positions respectively whereas Ag atoms occupy 4b ($\nicefrac{1}{2}$, $\nicefrac{1}{2}$, $\nicefrac{1}{2}$) and 4a (0, 0, 0) Wyckoff positions. The optimized lattice constant is 4.91 \AA{}. The dynamical stability of the structure was verified by computing the phonon dispersion (\emph{cf.} Supplemental Material). Fig. \protect\ref{fig:crystal_structure}\subref{subfig:crystal_and_BZ} shows the bulk Brillouin zone and its projection on to the equivalent (111) surface. This crystal structure has three-fold rotational symmetry ($\widetilde{C}_{3z}$) along the (111) direction, and there are four such equivalent directions. 

The band structure of Ag$_{2}$TmMg along high symmetry lines in the absence of SOC is shown in Fig. \protect\ref{fig:crystal_structure}\subref{subfig:bandstructure_nosoc}. In the neighborhood of Fermi level, along the $\Gamma$--X direction, a non-degenerate band touches a doubly degenerate band at around -0.31 eV. The $\Gamma$ point itself hosts a triply degenerate point at -0.16 eV as seen in the K -- $\Gamma$ -- L segment of the band structure. When projected onto the (111) surface, the TP observed at the $\Gamma$ point merges with the [111] axes, hence cannot be resolved experimentally. However, the other TPs, which are well separated from the $\Gamma$ point, can be observed on this easily cleavable (111) plane.

The phonon dispersion shows a peculiar crossing which is linearly dispersive as well as triply-degenerate. This suggests a bosonic Weyl quasiparticles.

\begin{figure*}
\begin{center}
\subfloat[] {\label{subfig:surfdos_l}\includegraphics[width=0.5\textwidth,valign=c]{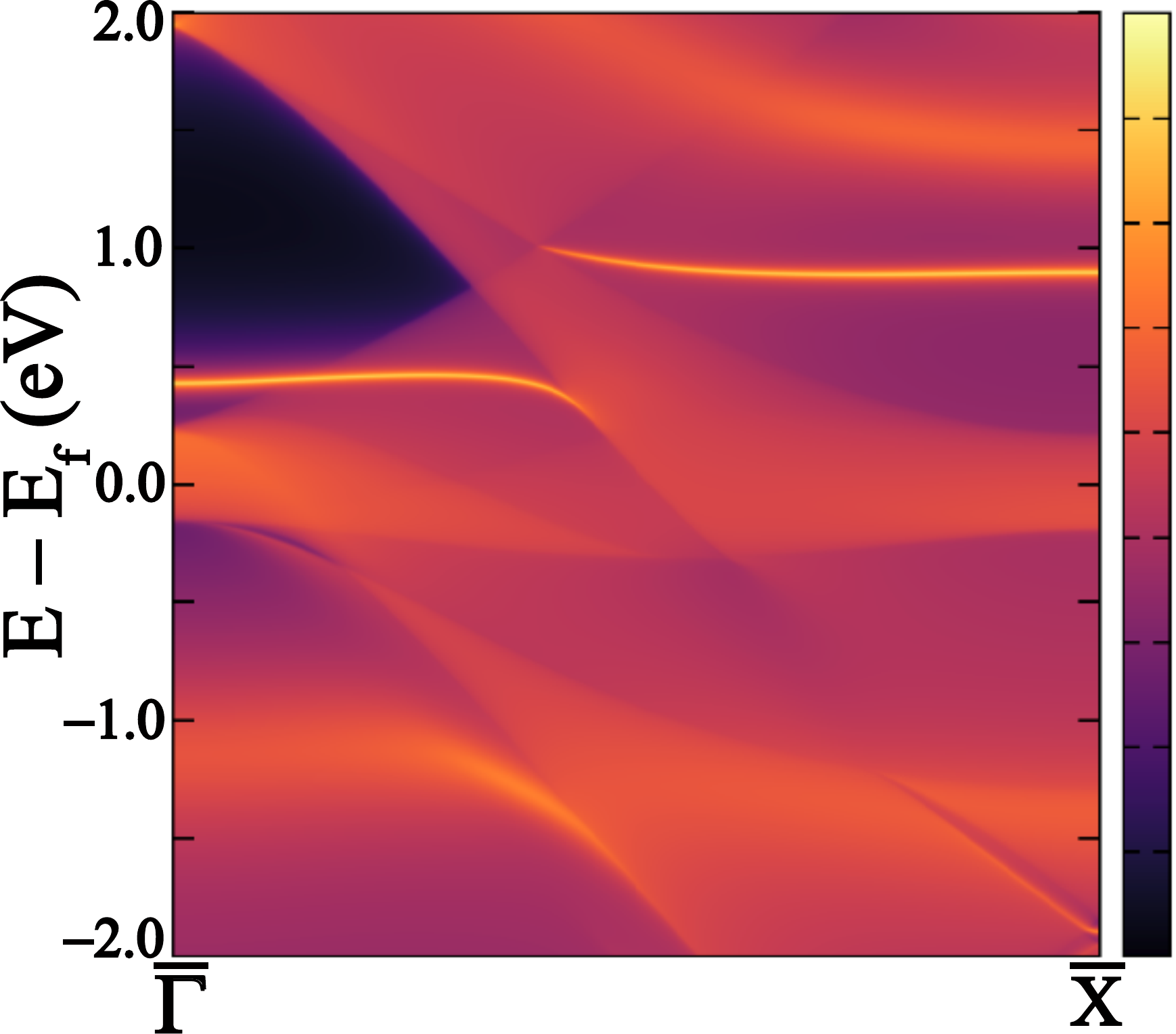}} 
\subfloat[] {\label{subfig:arc_l}\includegraphics[width=0.454\textwidth,valign=c]{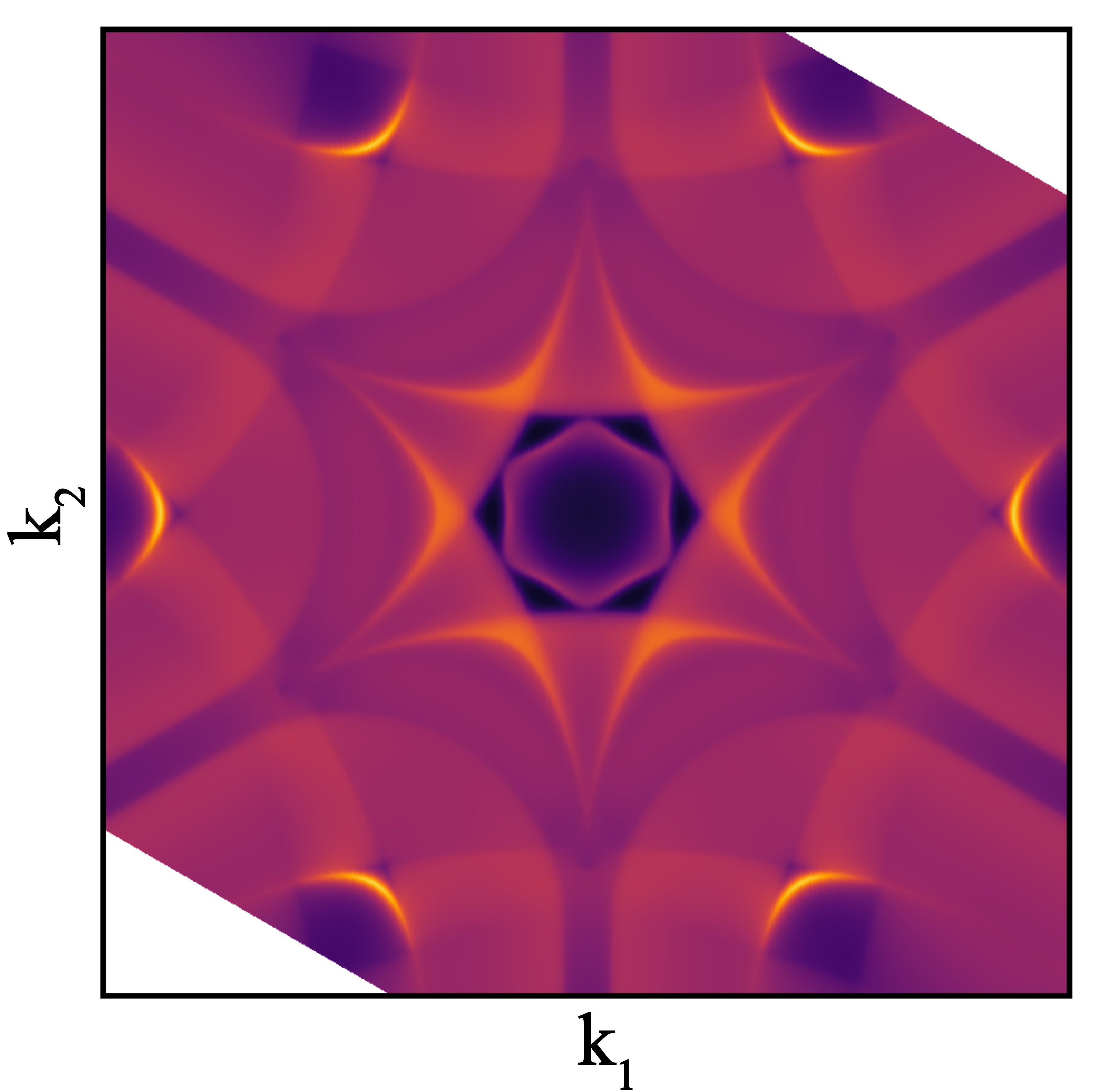}} \\
\subfloat[] {\label{subfig:surfdos_l_gamma}\includegraphics[width=0.5\textwidth,valign=c]{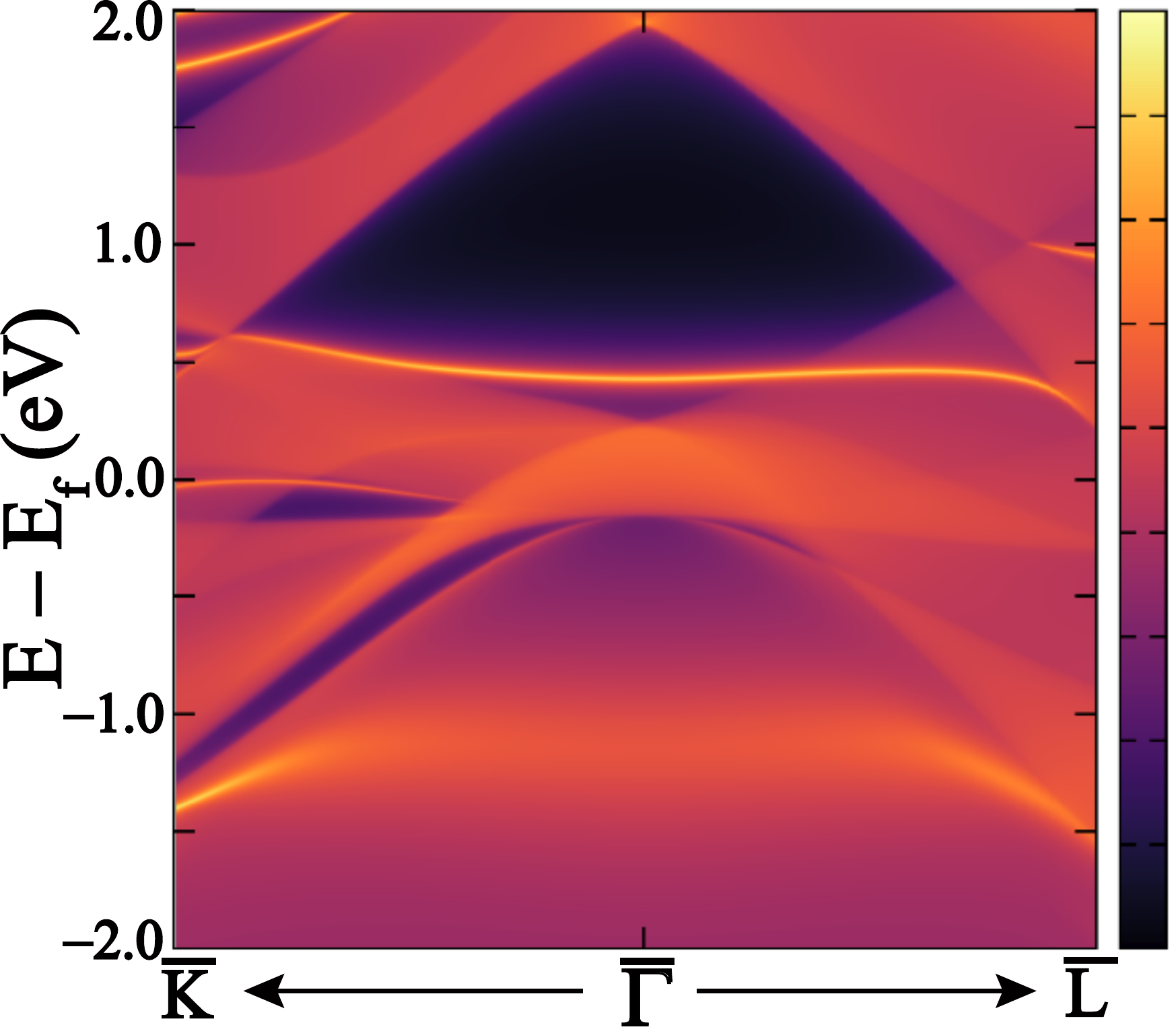}} 
\subfloat[] {\label{subfig:arc_l_gamma}\includegraphics[width=0.444\textwidth,valign=c]{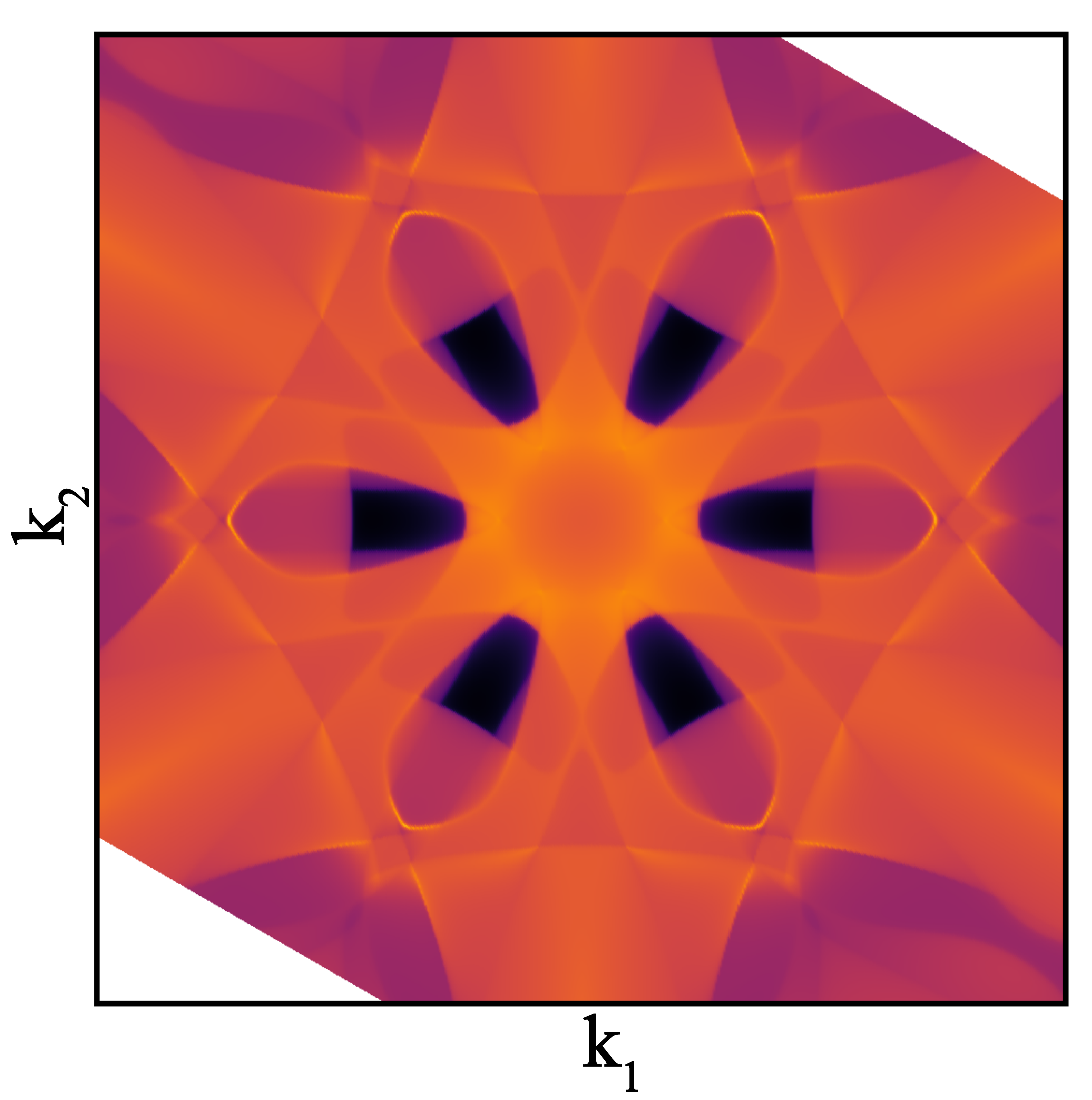}} \\
\end{center}
\caption{\label{fig:surface_fermi}  (a) The (111) Bloch spectral function depicting the surface states of Ag$_{2}$TmMg along the $\Gamma$ -- X segment. The dark yellow lines denote the surface contribution. (b) Fermi surface calculated with the chemical potential of  -0.31 eV.   (c) The (111) Bloch spectral function depicting the surface states of Ag$_{2}$TmMg near the $\Gamma$ point. The dark yellow lines denote the surface contribution. (d) Fermi surface calculated for the same with the chemical potential of  -0.16 eV.} 
\end{figure*}

\subsection{Bloch Spectral Function and Fermi Arcs}

A three-fold rotational symmetry ($\widetilde{C}_{3z}$) along (111) direction is a crucial requirement for the existence of TPs. Since there are four such equivalent (111) directions, the TPs will appear in four pairs in the Brillouin zone. With the inclusion of SOC, these TPs splits into two Dirac point of type-I and type-II along $\Gamma$-X direction as well as at $\Gamma$ point, as shown in the Fig. \protect\ref{fig:crystal_structure}\subref{subfig:bandstructure_soc}

The surface-projected density of states (or Bloch spectral function) is calculated by iterative Green's function method, by obtaining the tight-binding Hamiltonian from the maximally localized Wannier functions. This function simulated the experimental angle-resolved photoemission spectroscopy, giving valuable information about the surface states and nodal points. In Fig. \protect\ref{fig:surface_fermi}\subref{subfig:surfdos_l}, bright yellow lines indicate the contribution from the surface states, and the TP is seen at the -0.31 eV and the two-thirds of momenta along $\Gamma$-X direction. These states are seen to merge on the linearly dispersing bands and crossing the TP. The corresponding surface-projected density of states in the neighbourhood of $\bar{\Gamma}$ in the $\bar{K} - \bar{\Gamma} - \bar{L}$ direction is as shown in Fig. \protect\ref{fig:surface_fermi}\subref{subfig:surfdos_l_gamma}.  Slightly below the Fermi level, a TP is observed due to touching of non-linearly dispersing bands. This TP, however, cannot be resolved in surface measurements because of its location at the $\Gamma$ point. 

The TPs can also be characterized by the Fermi arcs on the surface Bloch spectrum. Since there are four equivalent (111) directions, the $\Gamma$ point projection coincides with one of the TPs on the (111) surface. Hence, three other TPs and their time-reversal counterparts will be seen in an hexagonal pattern on the (111) surface (\emph{cf} Fig. \protect \ref{fig:surface_fermi}\subref{subfig:arc_l}). The Fermi arcs emanating from a given TP connect to the neighboring TPs forming a flower-like hexagonal Fermi surface. Again, this pattern is due to the three-fold rotational symmetry and time-reversal symmetry along the diagonal of the primitive unit cell. 
The Fermi surface corresponding to the $\bar{\Gamma}$ in the $\bar{K} - \bar{\Gamma} - \bar{L}$ segment (\emph{cf} Fig. \protect\ref{fig:surface_fermi}\subref{subfig:arc_l_gamma}) shows a bright spot at the center connected by the arcs the nearby six TPs at the vertices of the hexagon.

\begin{figure}
\begin{center}
\includegraphics[width=0.5\textwidth,valign=c]{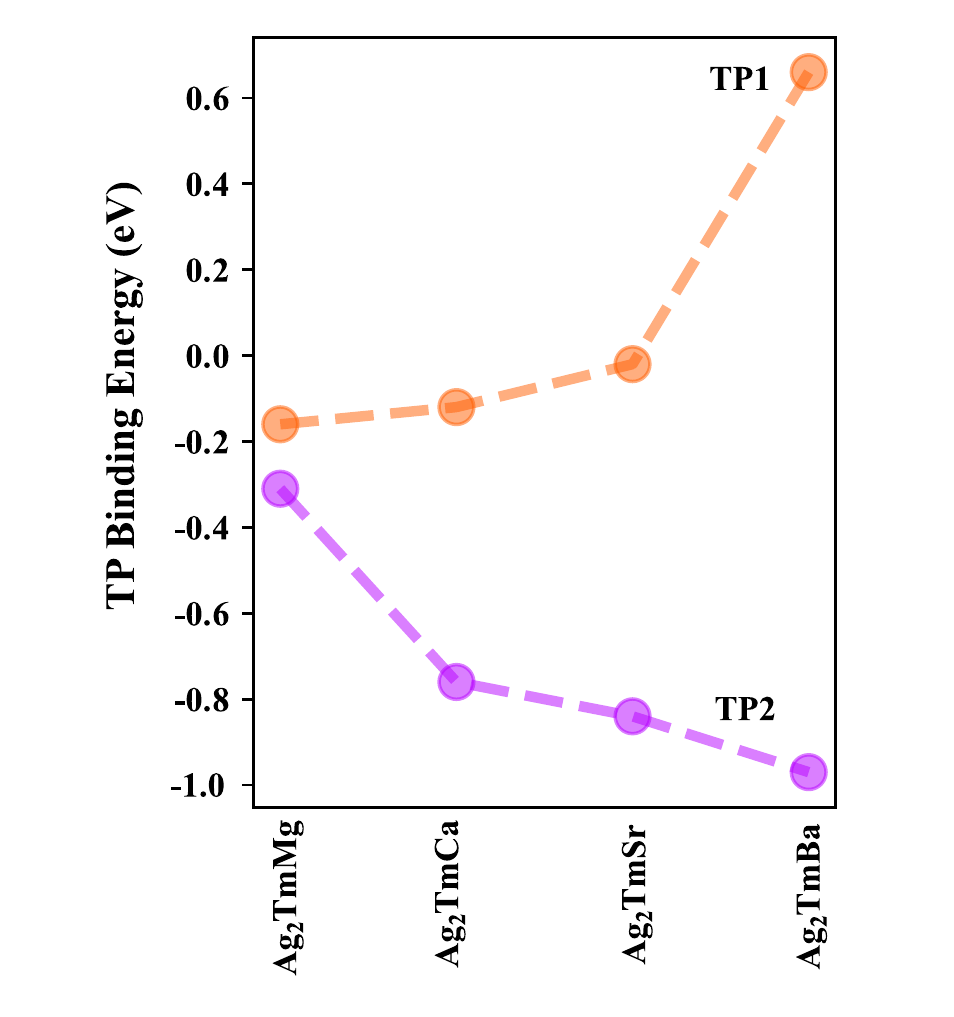} 
\end{center}
\caption{\label{fig:tp_variation}  The variations of binding energies of TPs with the selected compounds.} 
\end{figure}

In the case of Ag$_{2}$TmMg, we replaced the lighter element with the same group elements and observed its effect on the position of the two distinct TPs. Figure \ref{fig:tp_variation} shows the change in the location of TPs in the Brillouin zone for the different compounds. The TP1 (observed along the $\Gamma$ - X segment) moves down with increasing row of constituent elements, while TP2 moves upward. So, effectively we can tune the position of TPs in the appropriately alloyed compound.

\FloatBarrier

\section{Conclusion}
In conclusion, we have theoretically predicted several new materials which host peculiar triple-point fermions as quasiparticles. Also, they also show signature hexagonal Fermi arcs connecting the vertices to the center and the nearest vertex. This set of materials and the slab were particularly chosen to maximize the number of TPs and easy experimental preparation. Since all the six TPs are well separated from each other and are away from the Brillouin zone center, it offers easy detection using surface sensitive spectroscopic methods.  The two distinct TPs observed shift anti-parallel with the change in the row of the constituent lighter atom. This study can guide the experimental observation of triple-point fermions, based on the location of Fermi arcs. The position of the TPs can also be tuned with an appropriate alloying. Additionally, not only electronic bands show triply-degenerate crossings, but also the phonon bands exhibit this characteristic behaviour. The TP fermions cannot be characterized by the conventional topological invariants, elucidating the complexity of this new class of topological materials.

\FloatBarrier

\section{Suplemental Material}
The phonon dispersion and band structures of materials are given in the supplemental material.

\acknowledgements
The authors thank Dr. Swaminathan Venkataraman for valuable discussions. R.S. acknowledges Science and Engineering Research Board, DST, India for a fellowship (PDF/2015/000466).  This work is partly supported by the U.S. Army Contract FA5209-16-P-0090. We also acknowledge Materials Research Center and Supercomputer Education and Research Center, Indian Institute of Science for providing computational facilities. 

\bibliography{references}
\end{document}


\maketitle
\newpage

\section{Phonon}
  The phonon dispersion of all the compounds studied without spin-orbit coupling is given in Figure

\begin{figure*}[htb]
 \subfloat[Pd$_{2}$DyIn]{\includegraphics[width=0.55\columnwidth]{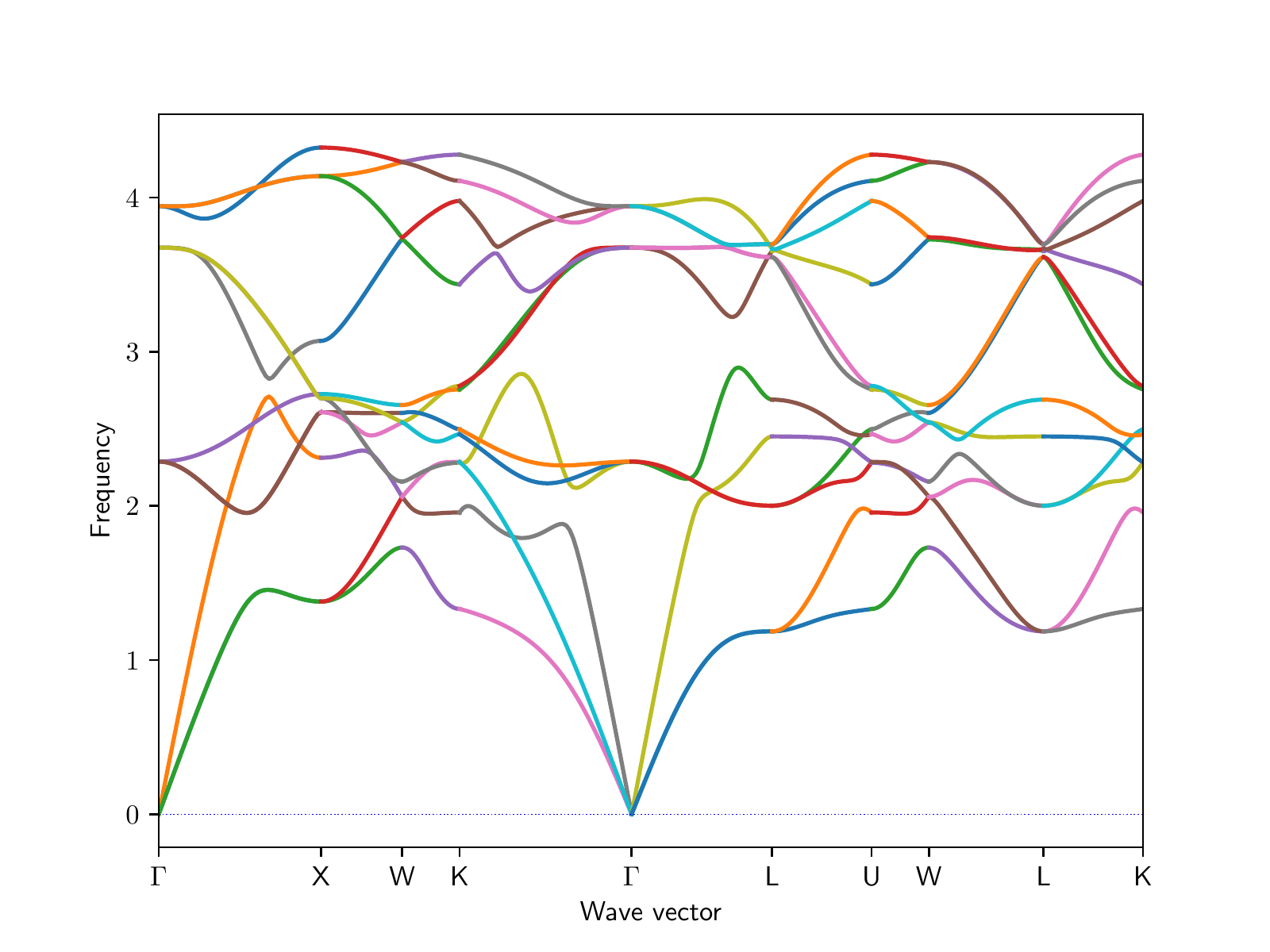}} 
 \subfloat[Hg$_{2}$LaMg]{\includegraphics[width=0.55\columnwidth]{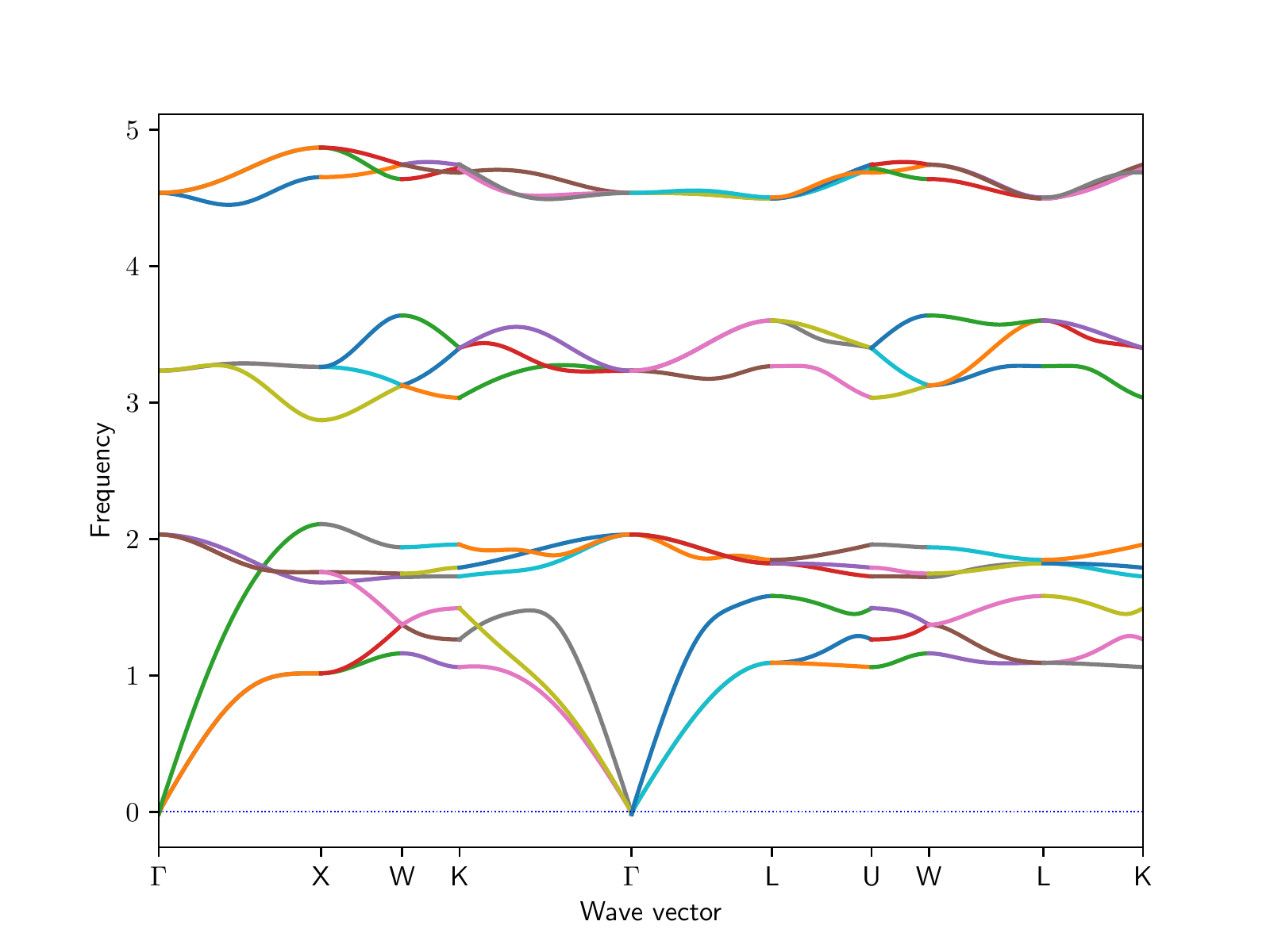}} \\
 \subfloat[Pd$_{2}$ErIn]{\includegraphics[width=0.55\columnwidth]{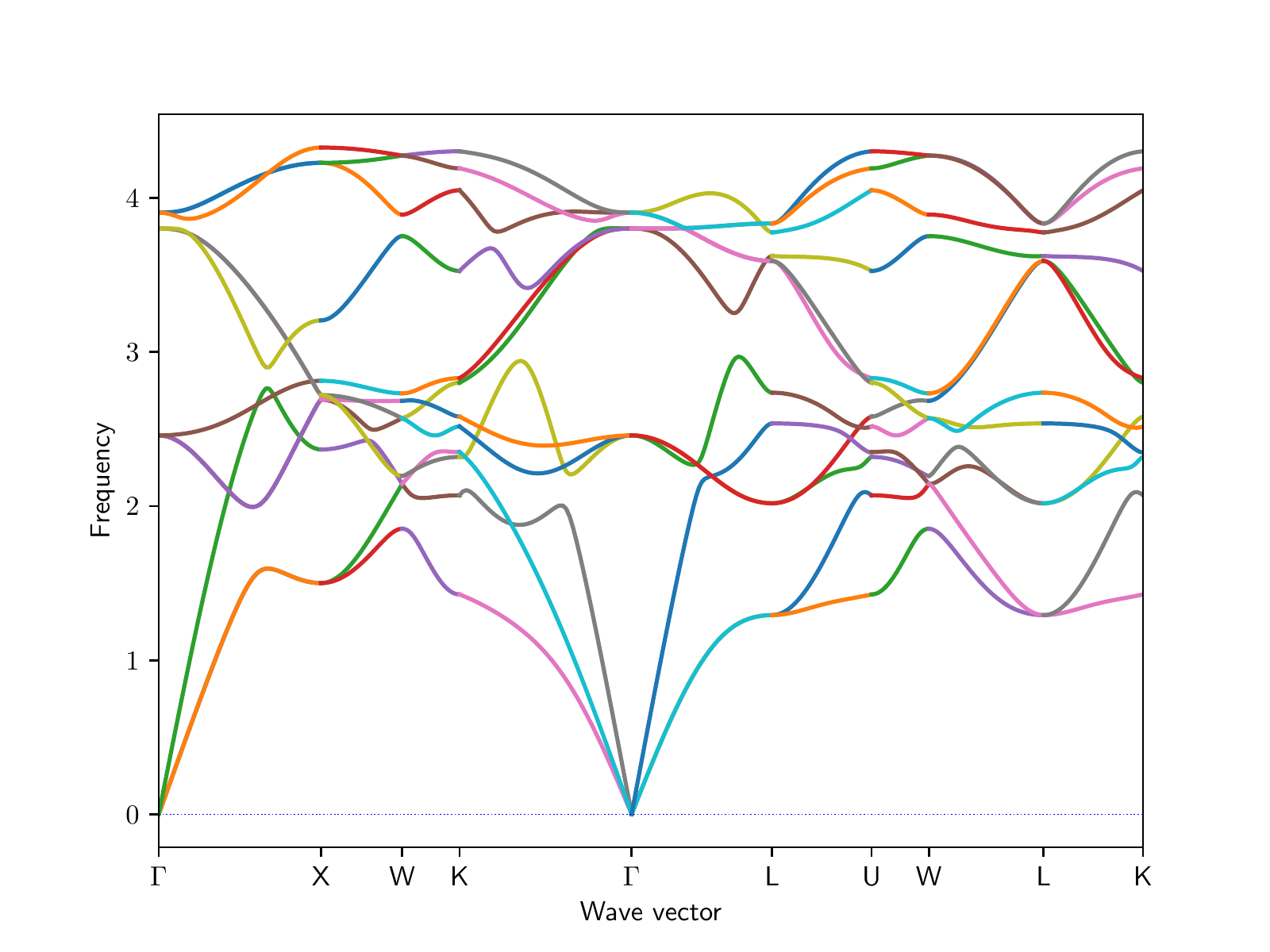}} 
 \subfloat[Hg$_{2}$PrAg]{\includegraphics[width=0.55\columnwidth]{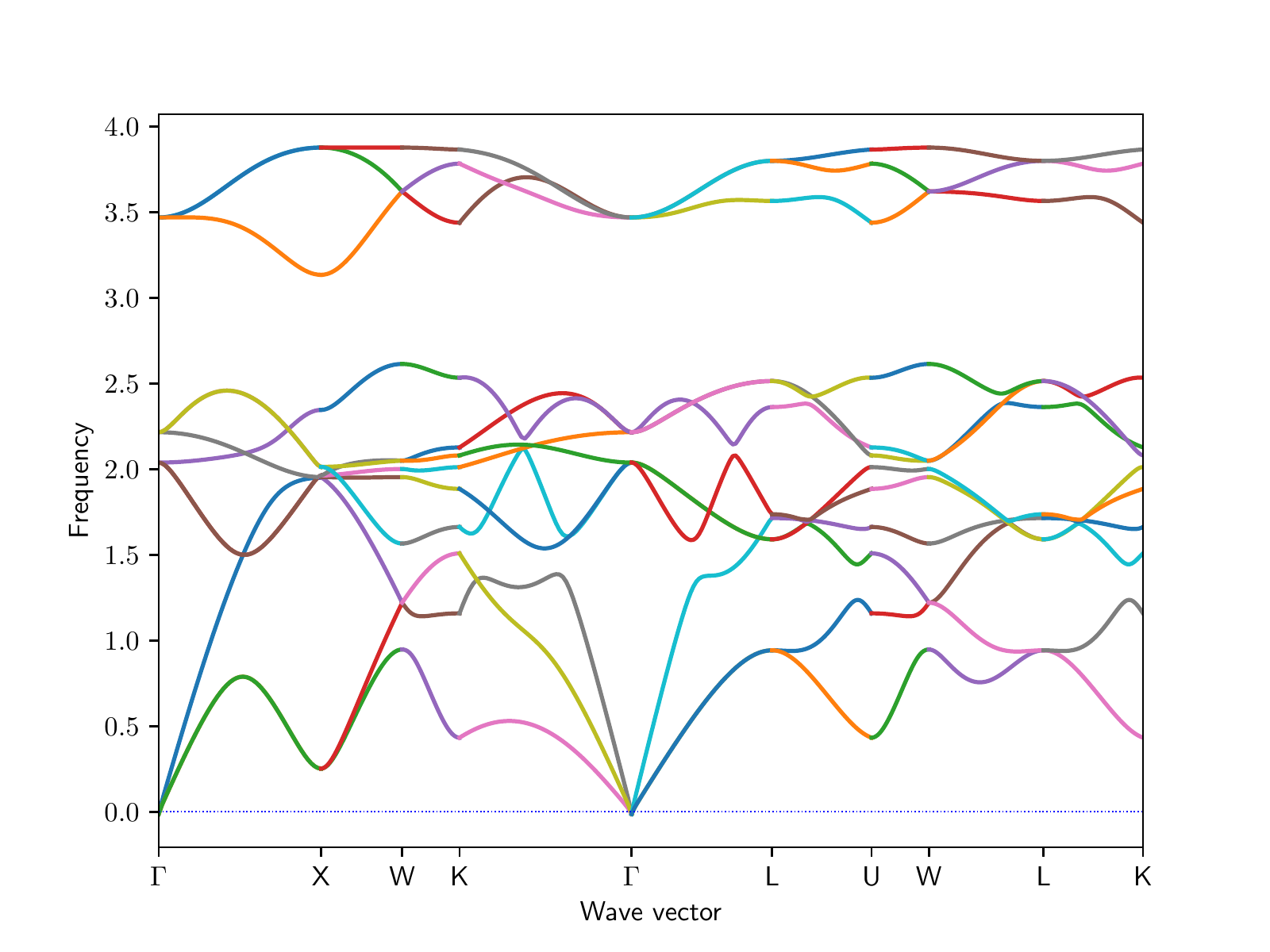}} \\
\end{figure*}

\begin{figure*}[htb]
 \ContinuedFloat
 \subfloat[Ag$_{2}$TbAl]{\includegraphics[width=0.55\columnwidth]{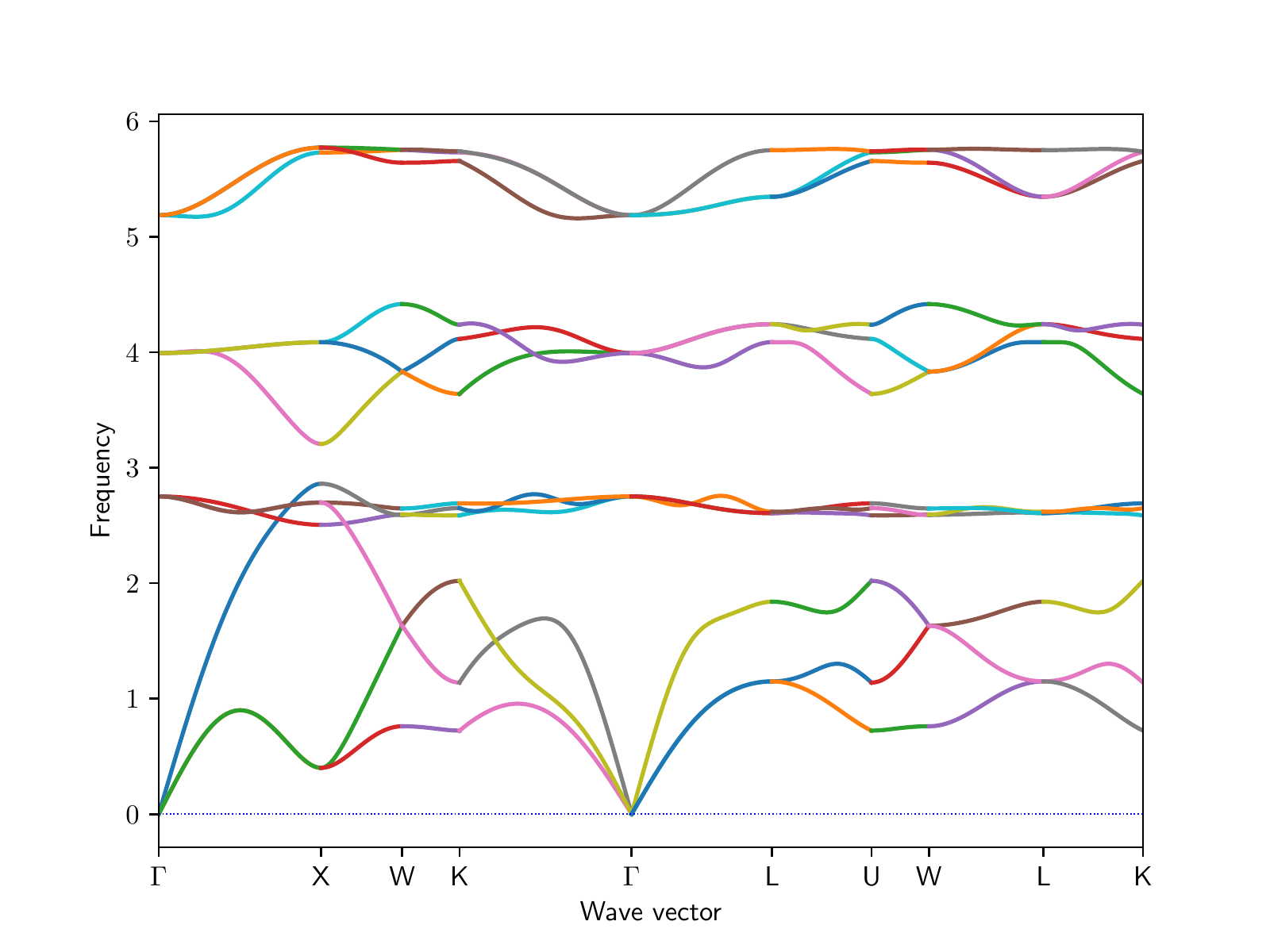}} 
 \subfloat[Ag$_{2}$YMg] {\includegraphics[width=0.55\columnwidth]{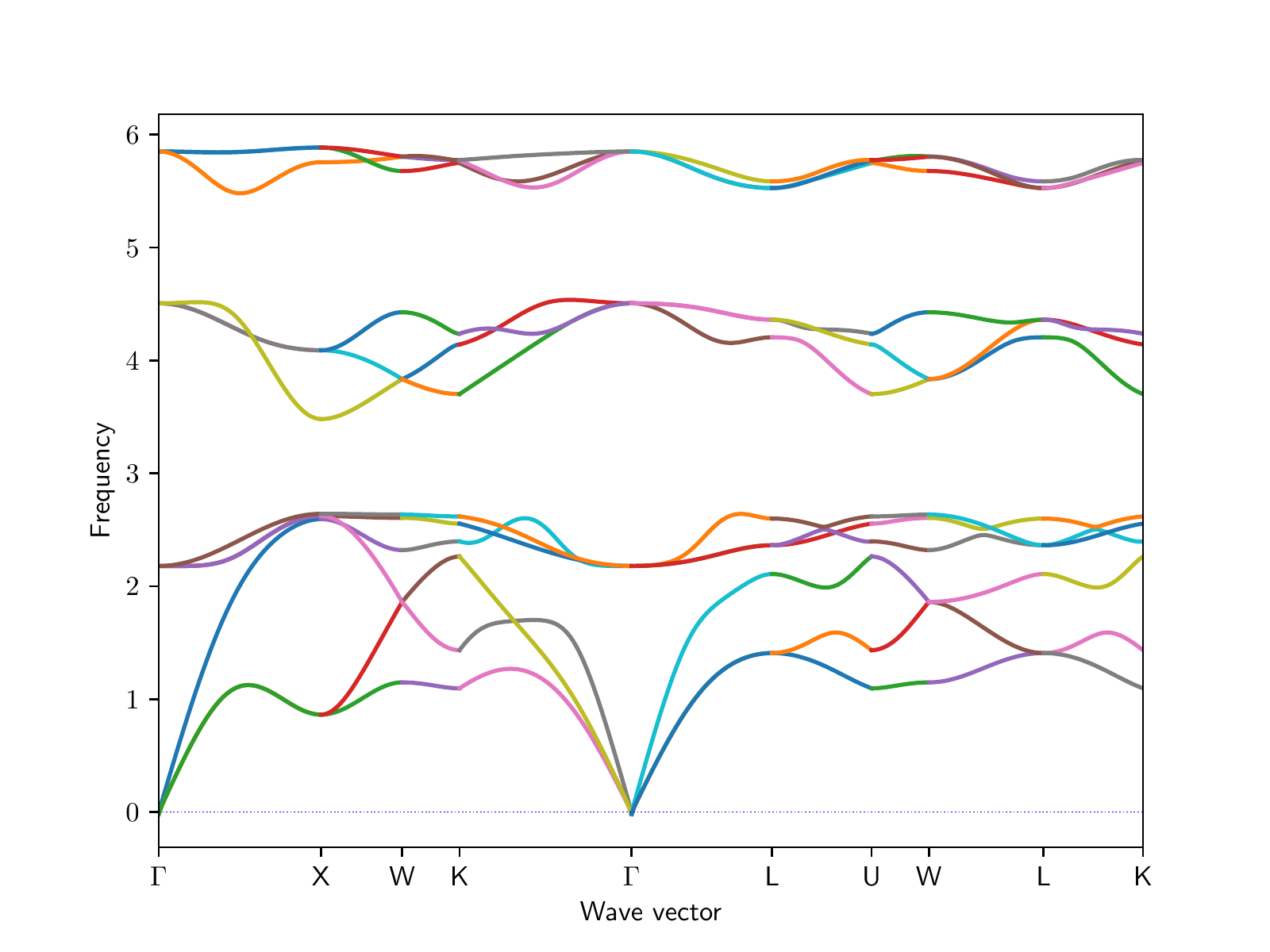}} \\
 \subfloat[Au$_{2}$DyMg]{\includegraphics[width=0.55\columnwidth]{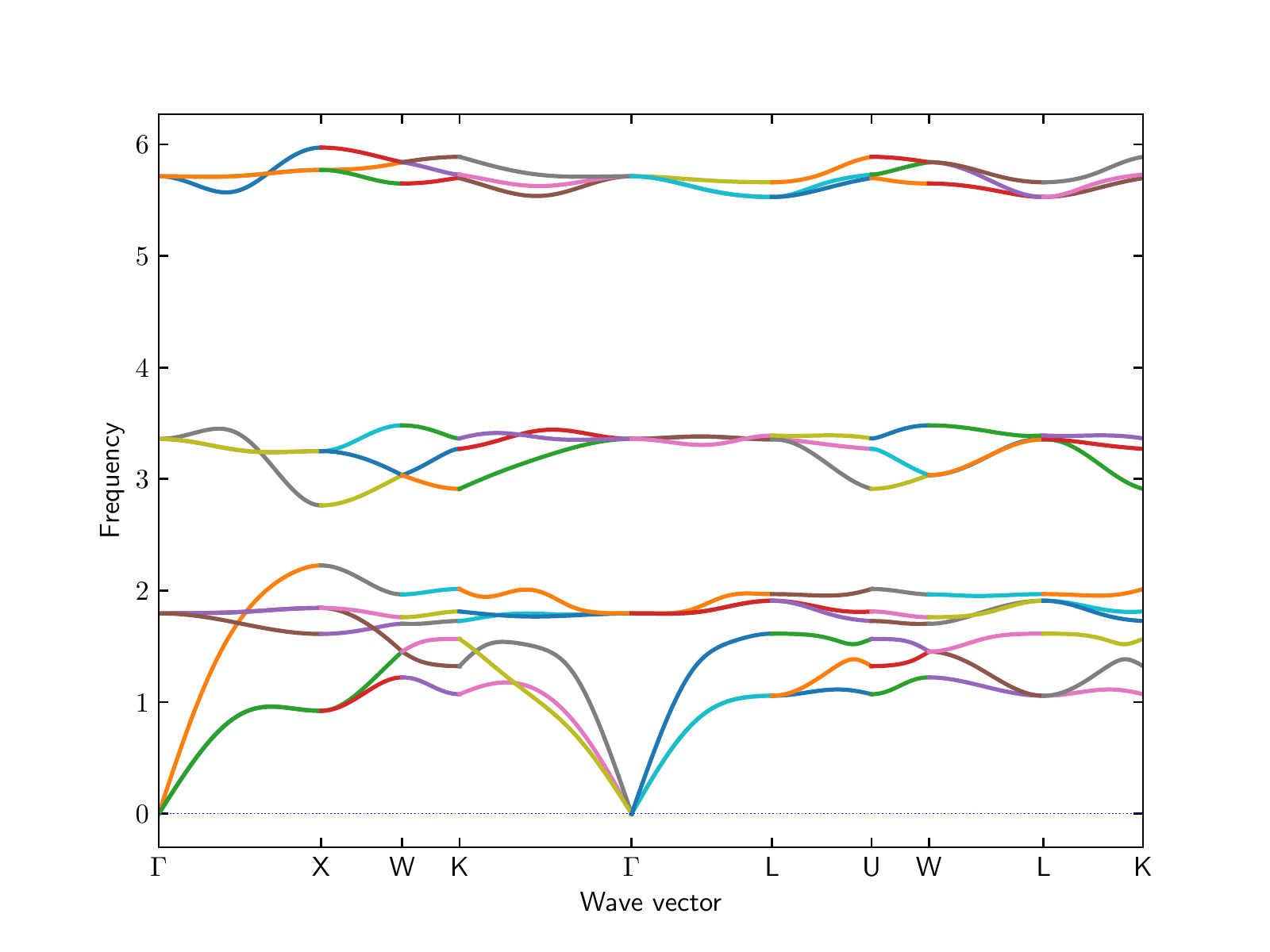}}
 \subfloat[Ag$_{2}$PrIn]{\includegraphics[width=0.55\columnwidth]{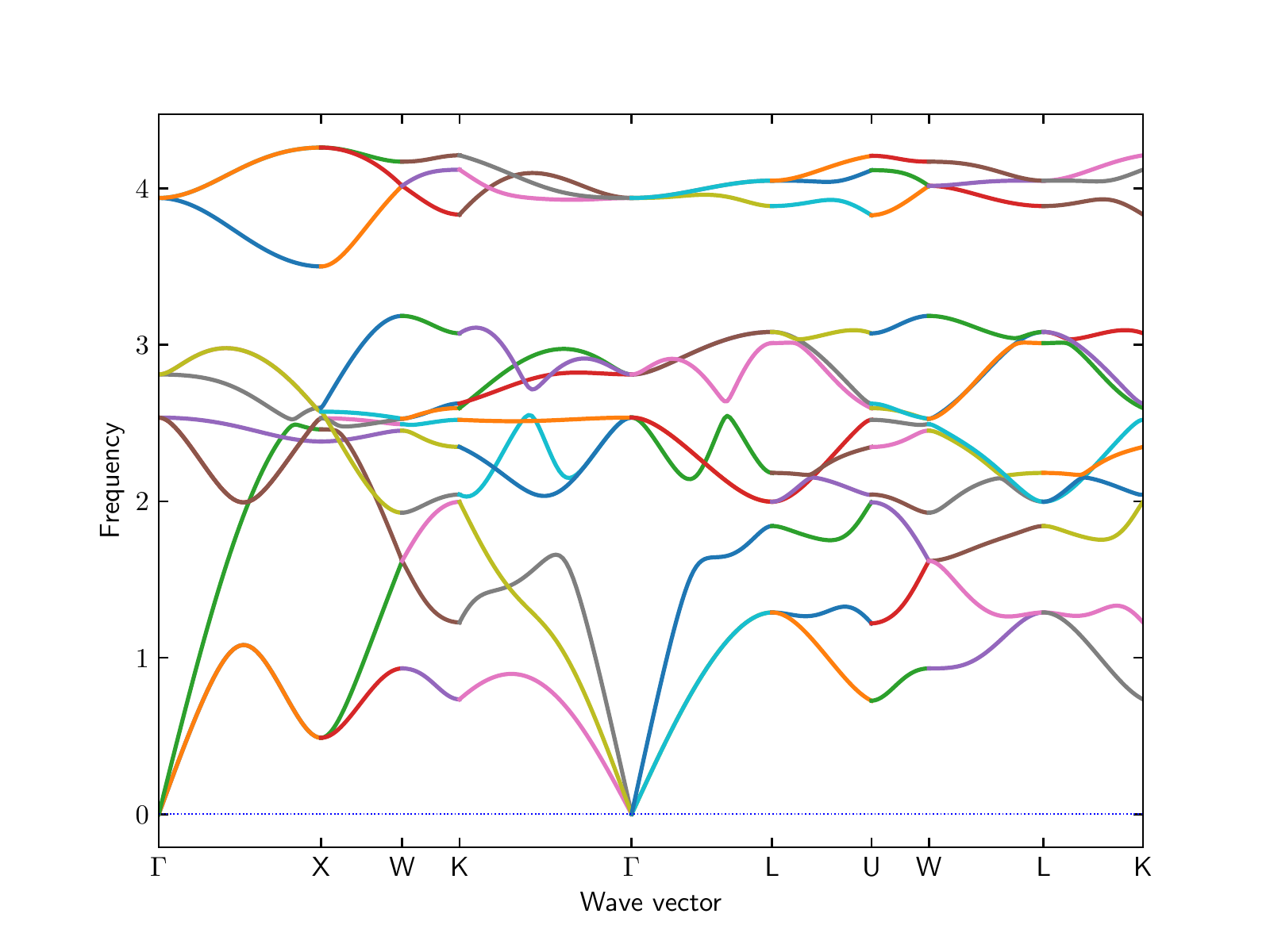}} \\
 \subfloat[Ag$_{2}$PrMg]{\includegraphics[width=0.55\columnwidth]{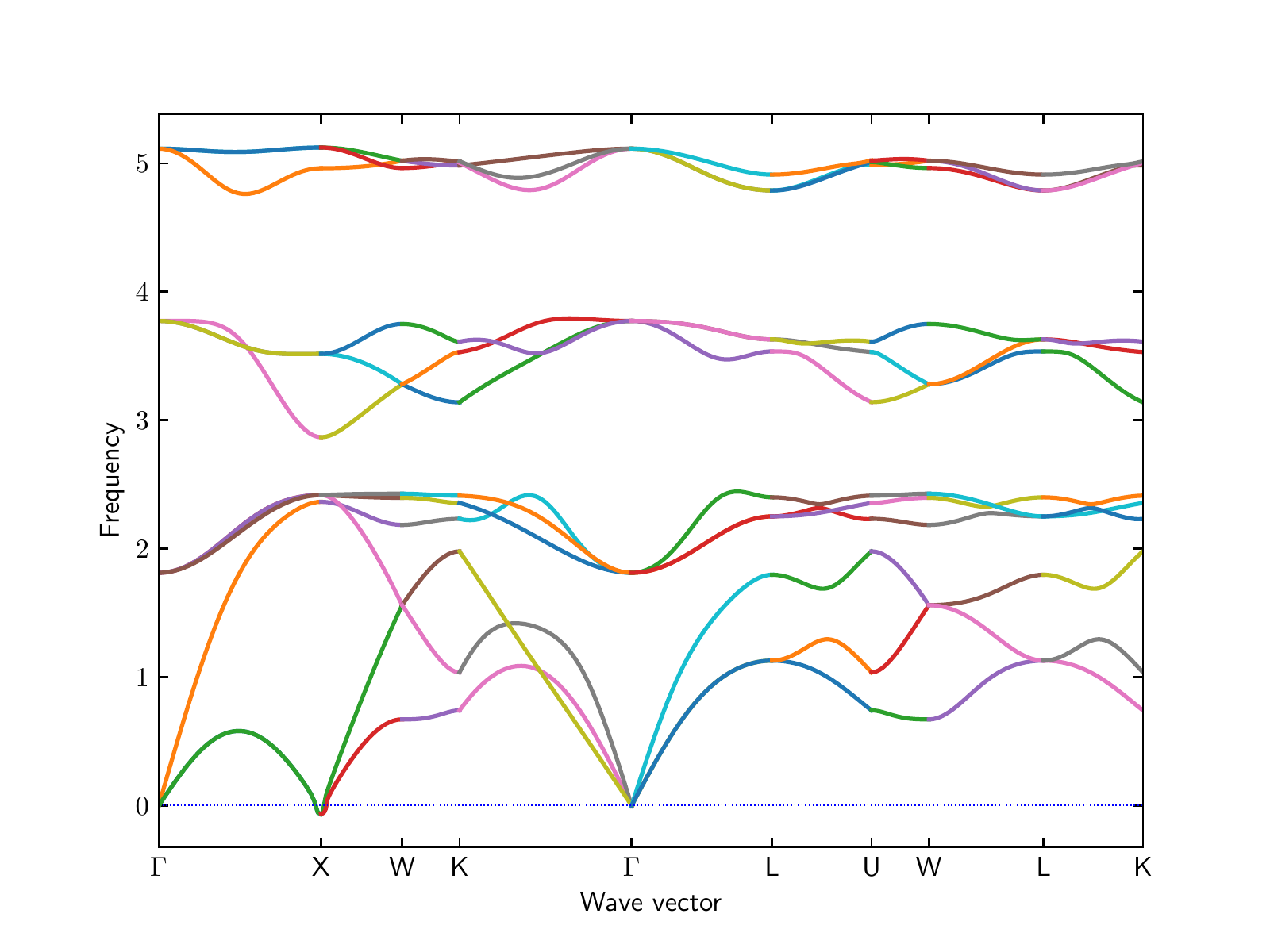}}
 \subfloat[Ag$_{2}$TbIn]{\includegraphics[width=0.55\columnwidth]{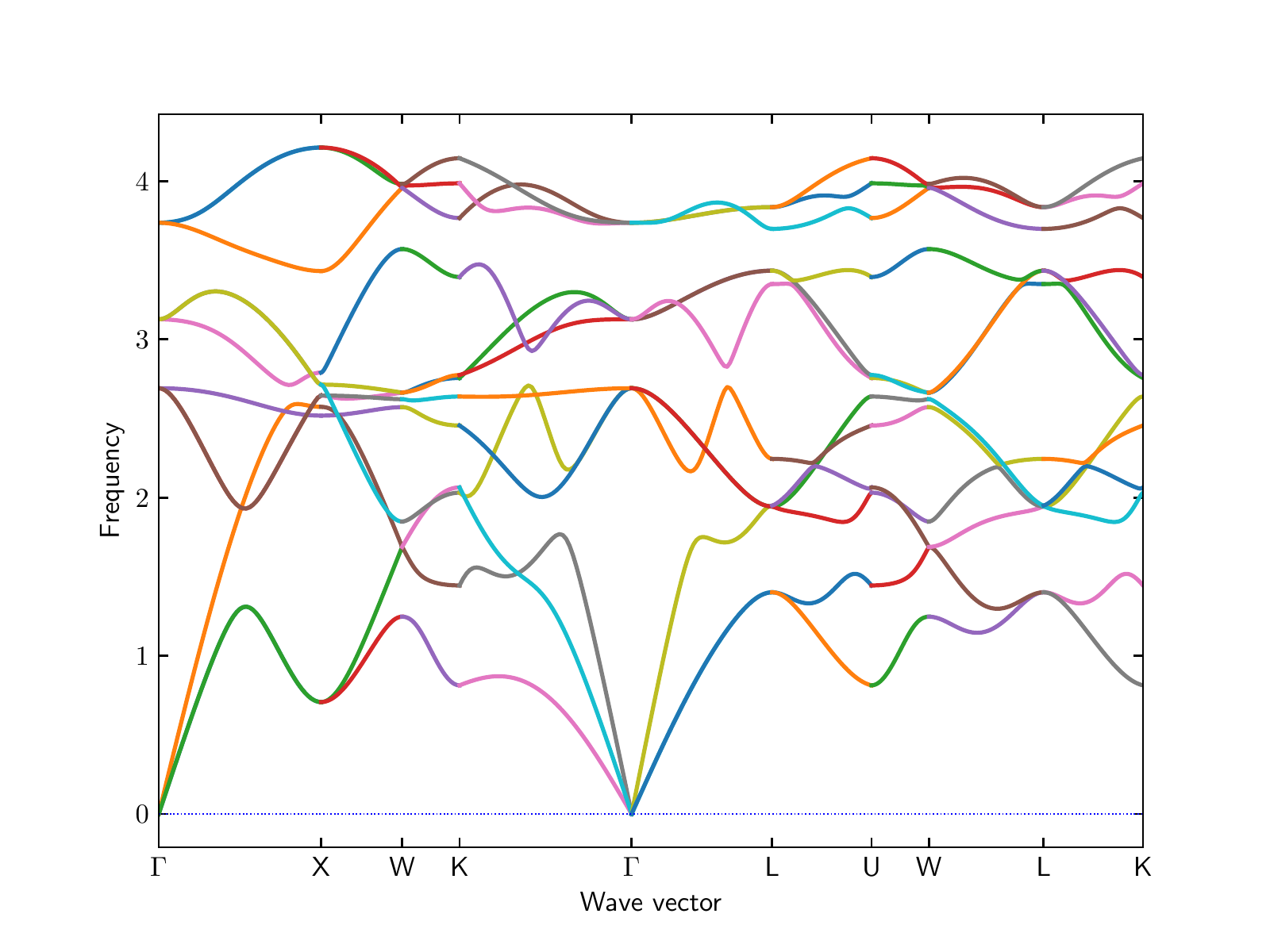}} \\
 \caption{The phonon band structure of all full Heuslers compounds}
 \label{fig:Phonon1}
\end{figure*}

\begin{figure*}[htb]
 \subfloat[Au$_{2}$DyCd]{\includegraphics[width=0.55\columnwidth]{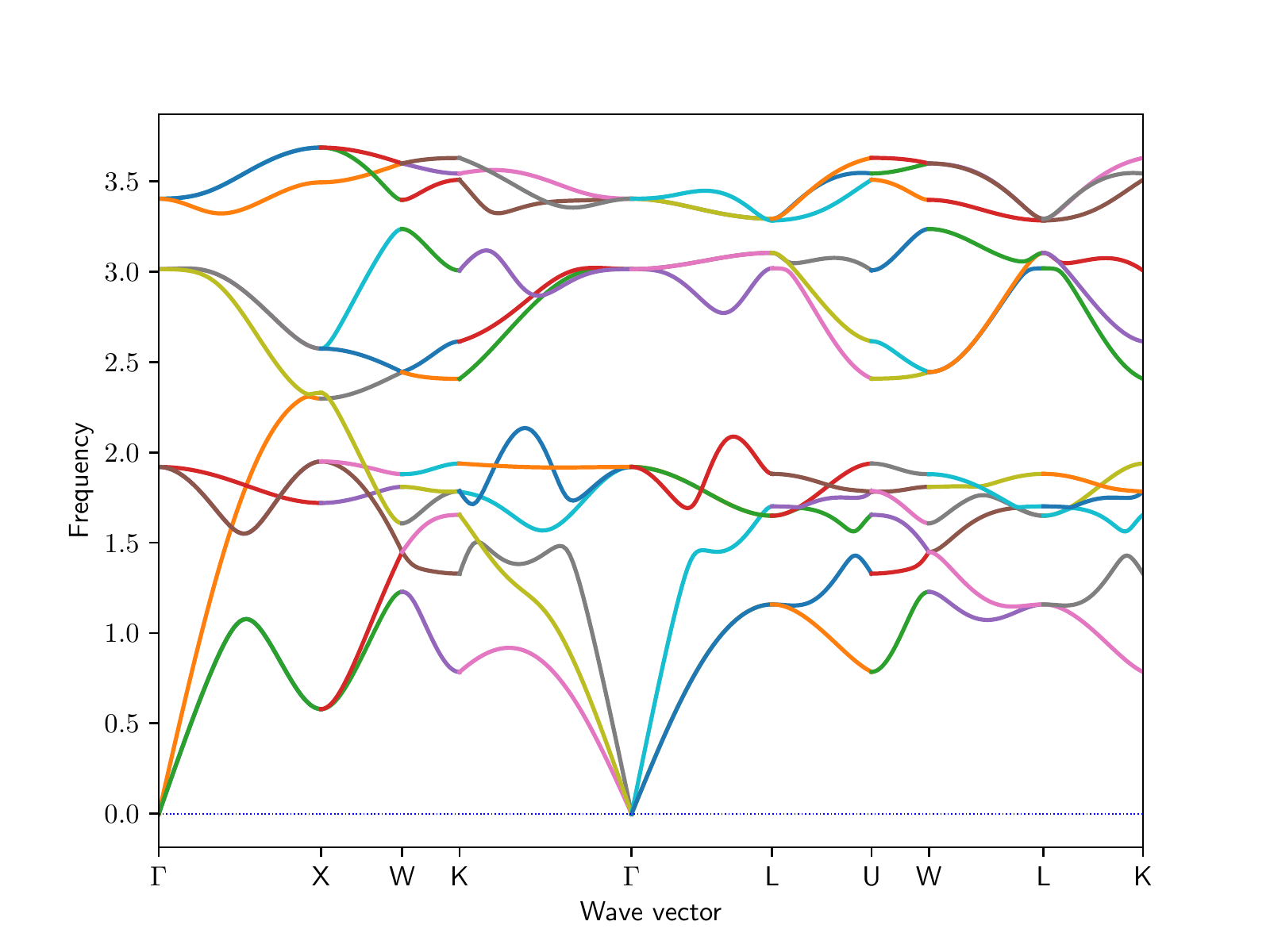}}
 \subfloat[Pd$_{2}$ScGa]{\includegraphics[width=0.55\columnwidth]{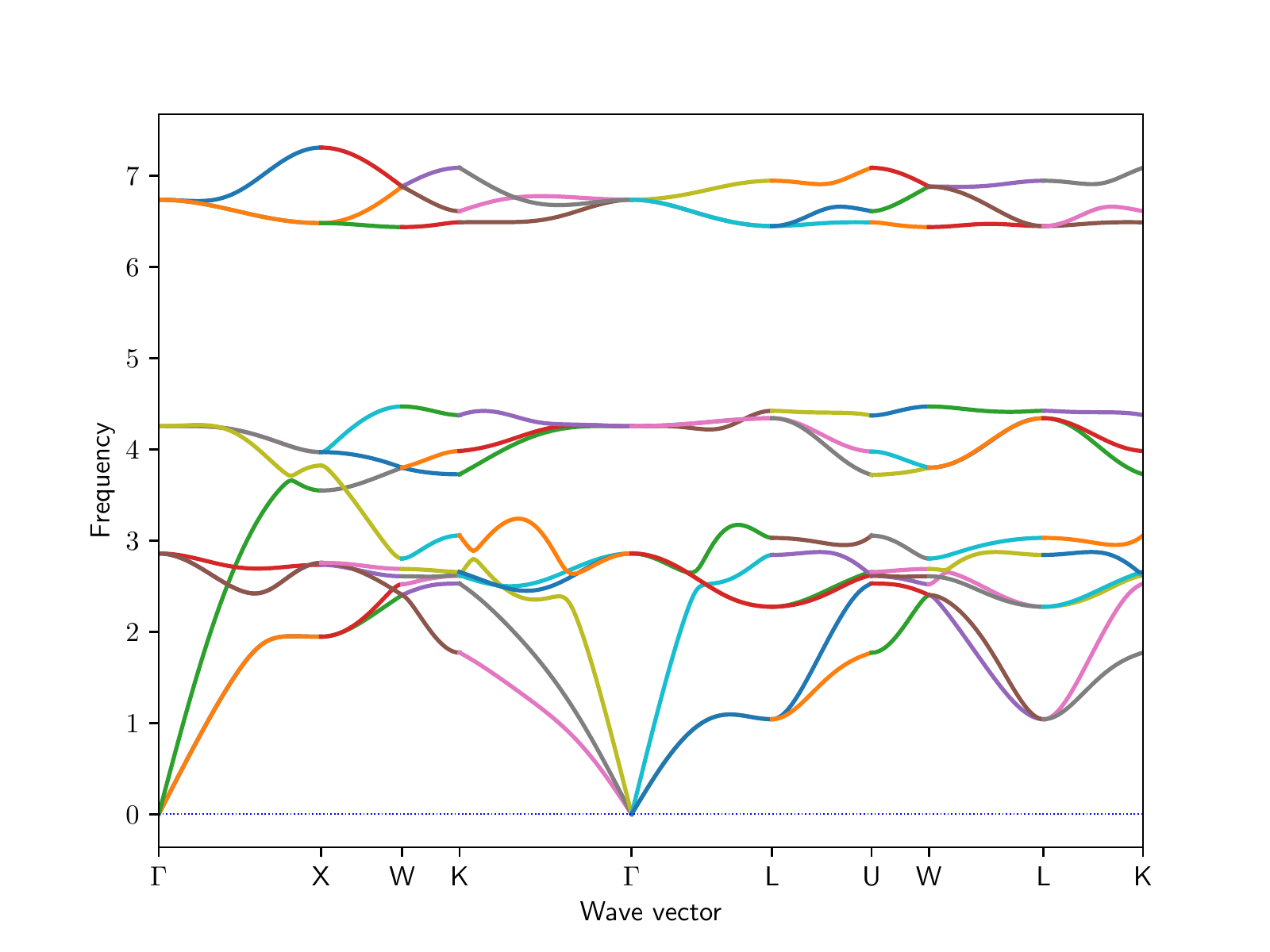}} \\ 
 \subfloat[Ag$_{2}$ScIn]{\includegraphics[width=0.55\columnwidth]{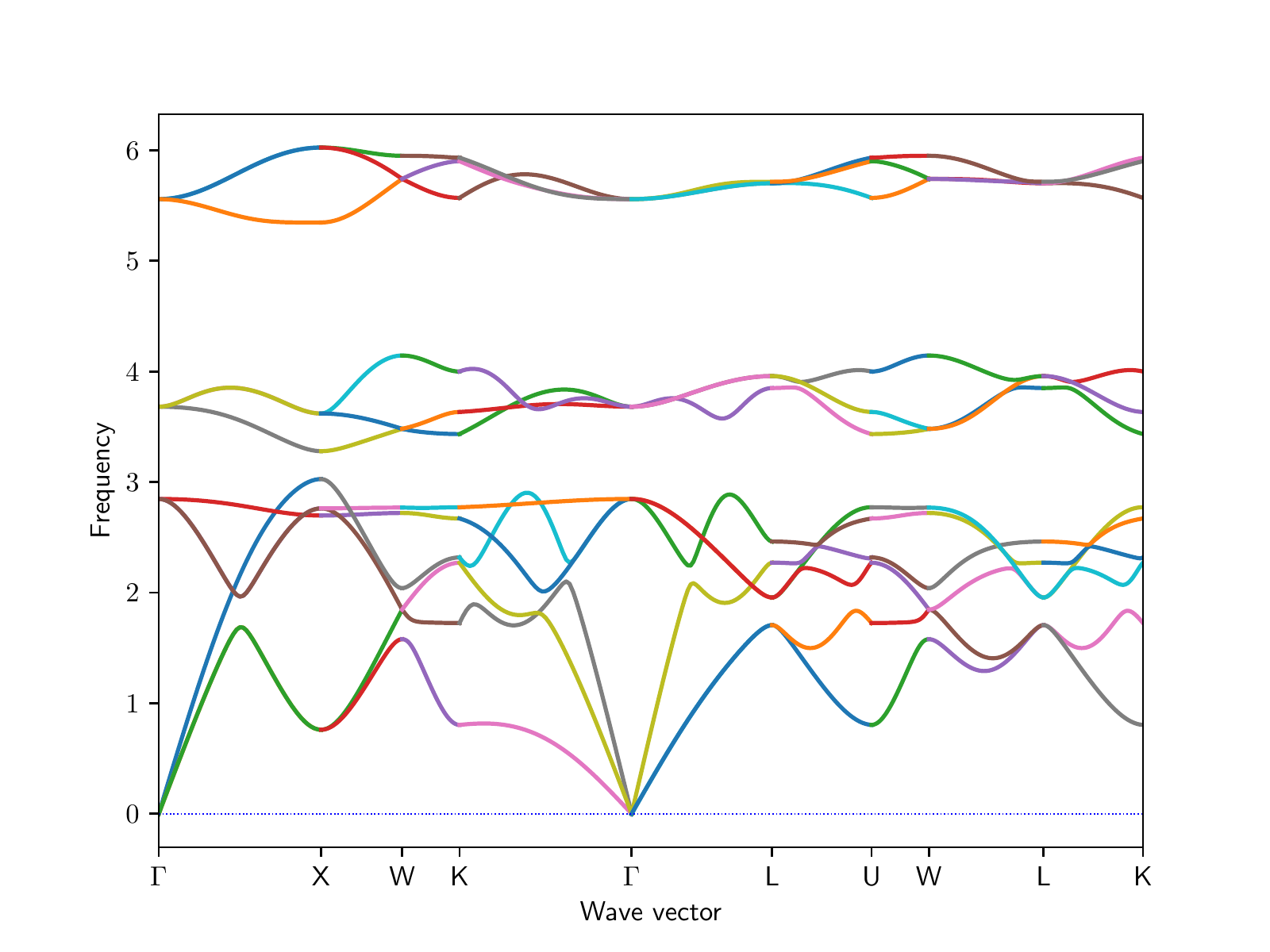}}
 \subfloat[Au$_{2}$PrIn]{\includegraphics[width=0.55\columnwidth]{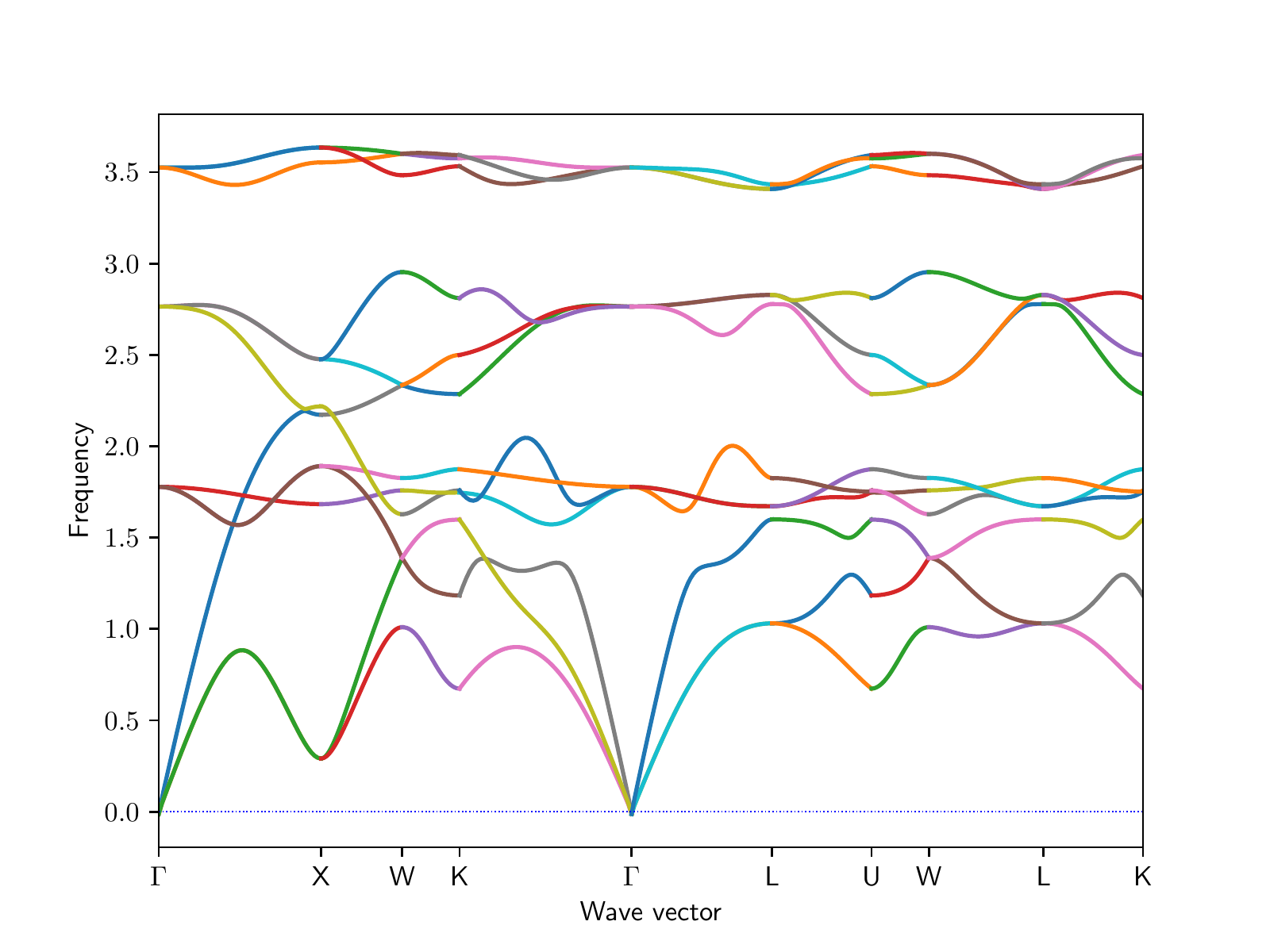}} \\
 \subfloat[Rh$_{2}$ZrAl]{\includegraphics[width=0.55\columnwidth]{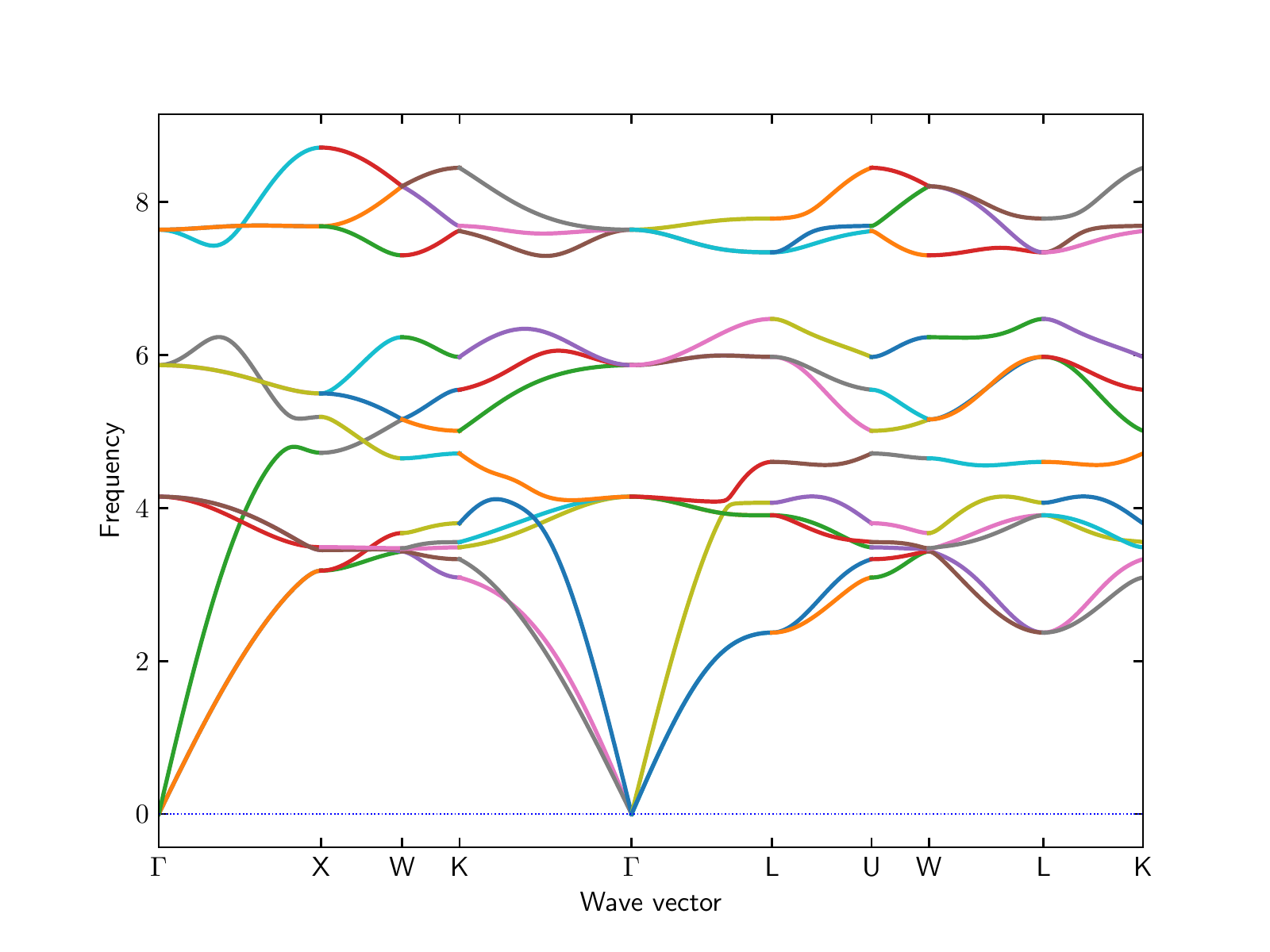}}
 \subfloat[Hg$_{2}$NaSm]{\includegraphics[width=0.55\columnwidth]{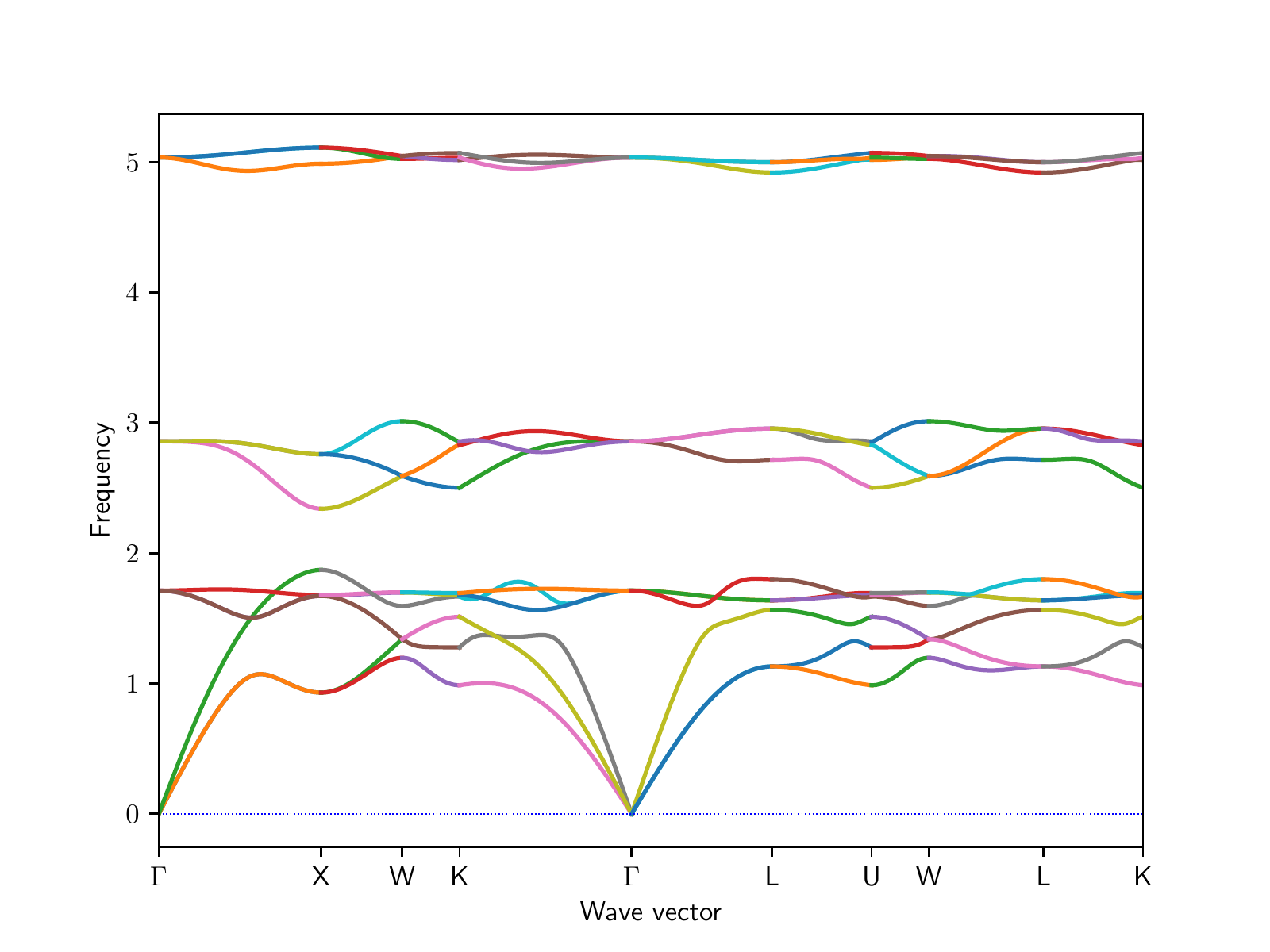}} \\ 
\end{figure*}

\begin{figure*}[htb]
 \ContinuedFloat
 \subfloat[Ag$_{2}$SmIn]{\includegraphics[width=0.55\columnwidth]{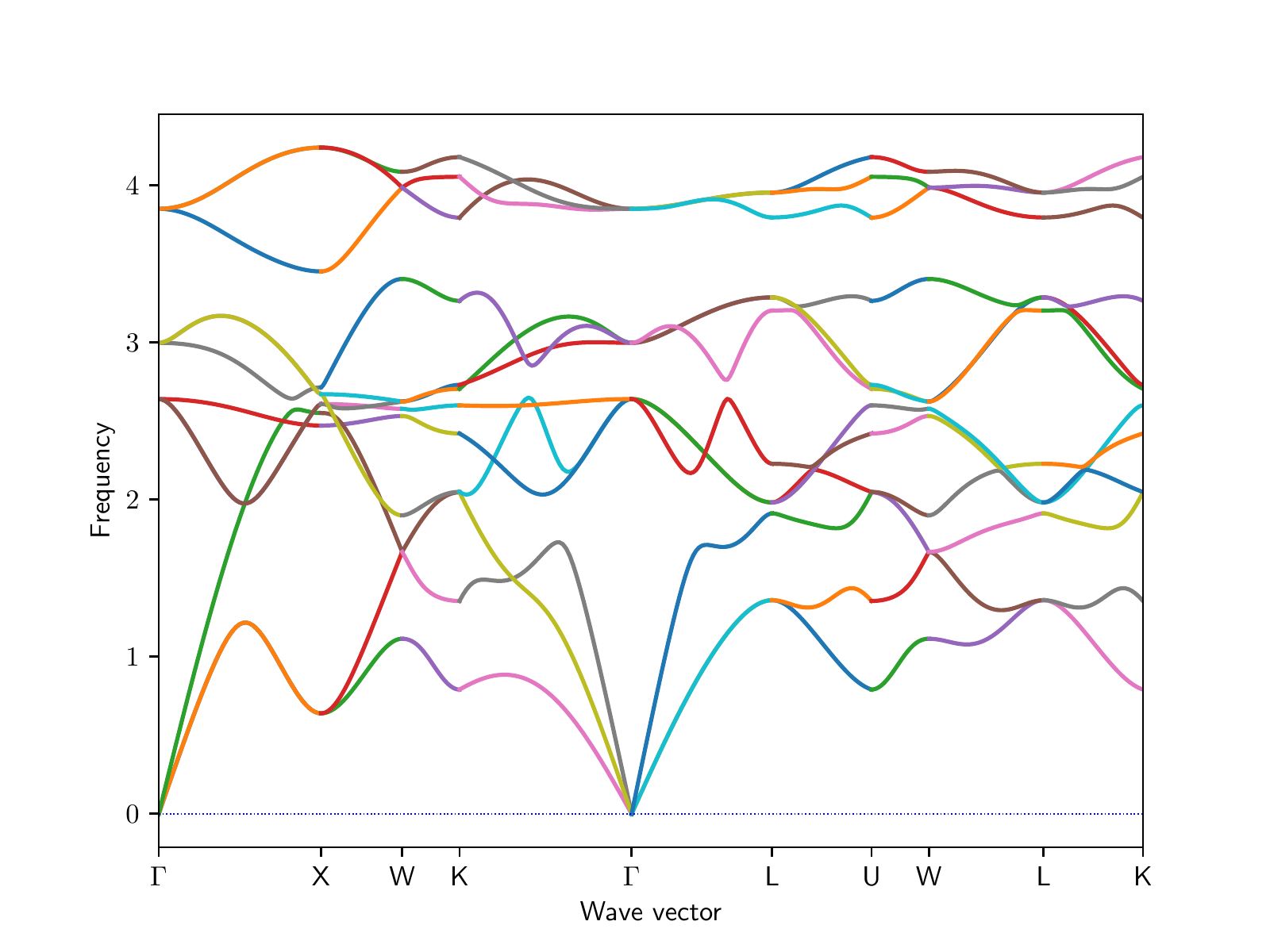}}
 \subfloat[Ag$_{2}$TmCa]{\includegraphics[width=0.55\columnwidth]{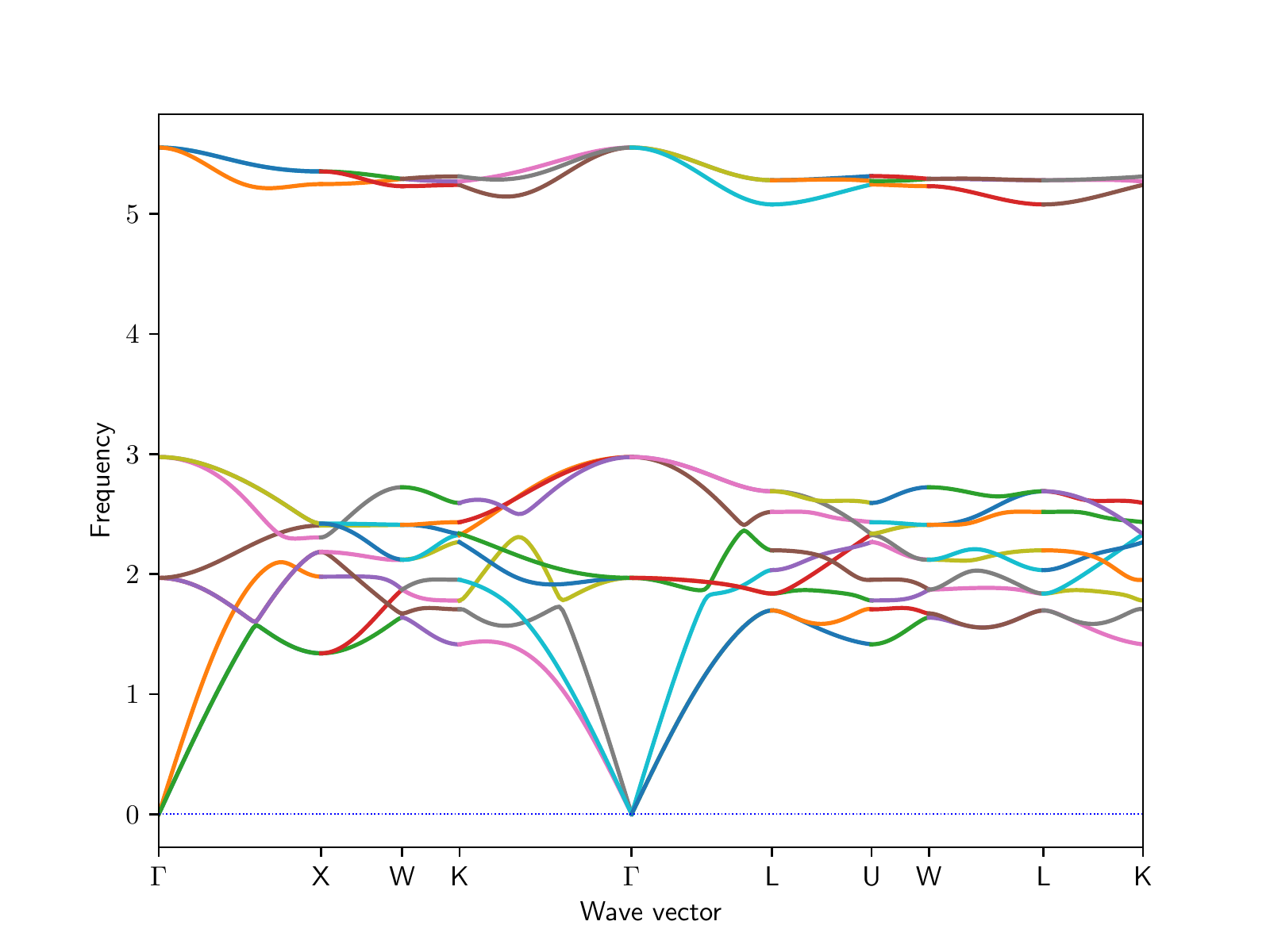}} \\
 \subfloat[Ag$_{2}$TmSr]{\includegraphics[width=0.55\columnwidth]{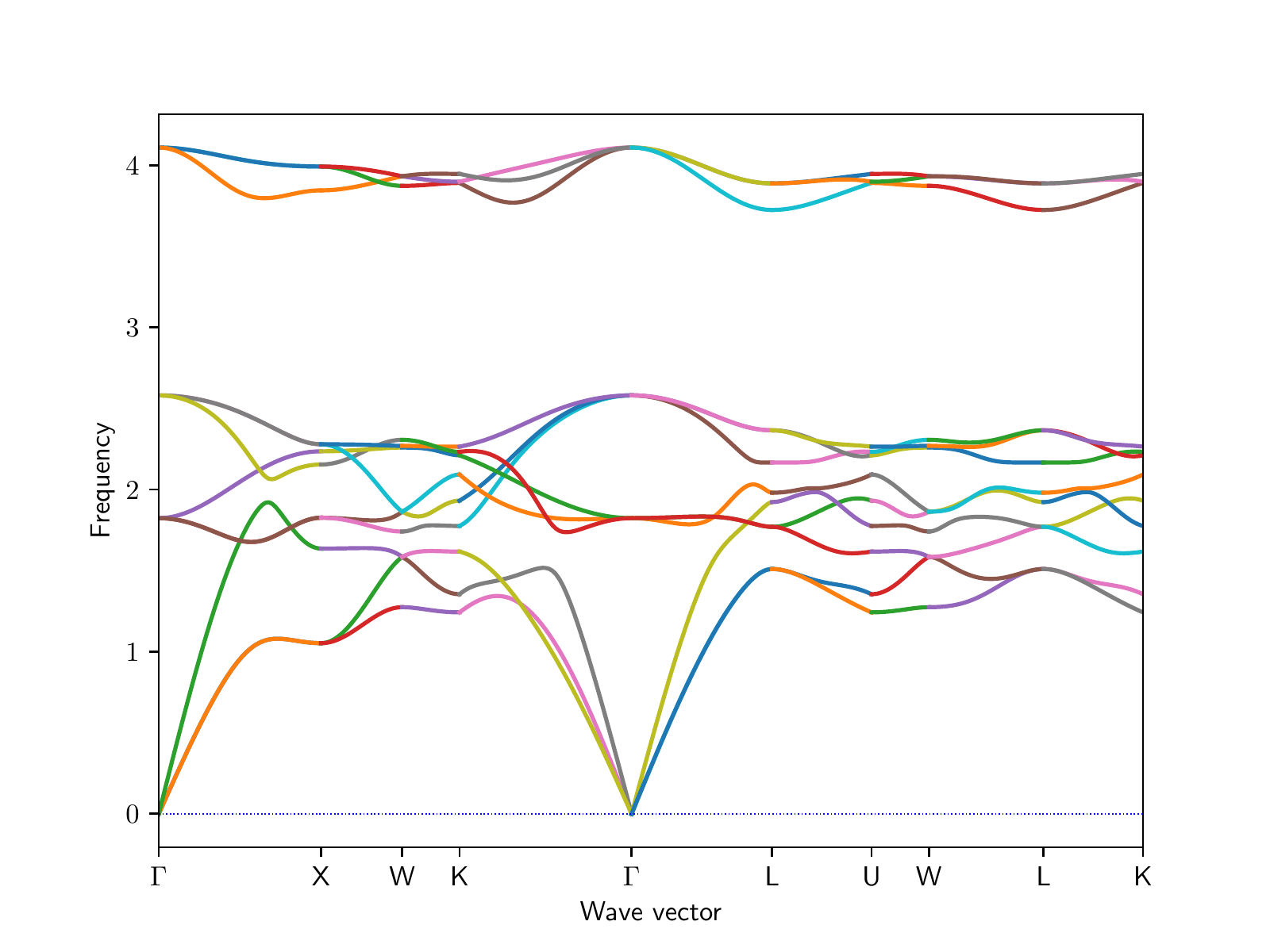}}
 \subfloat[Ag$_{2}$TmBa]{\includegraphics[width=0.55\columnwidth]{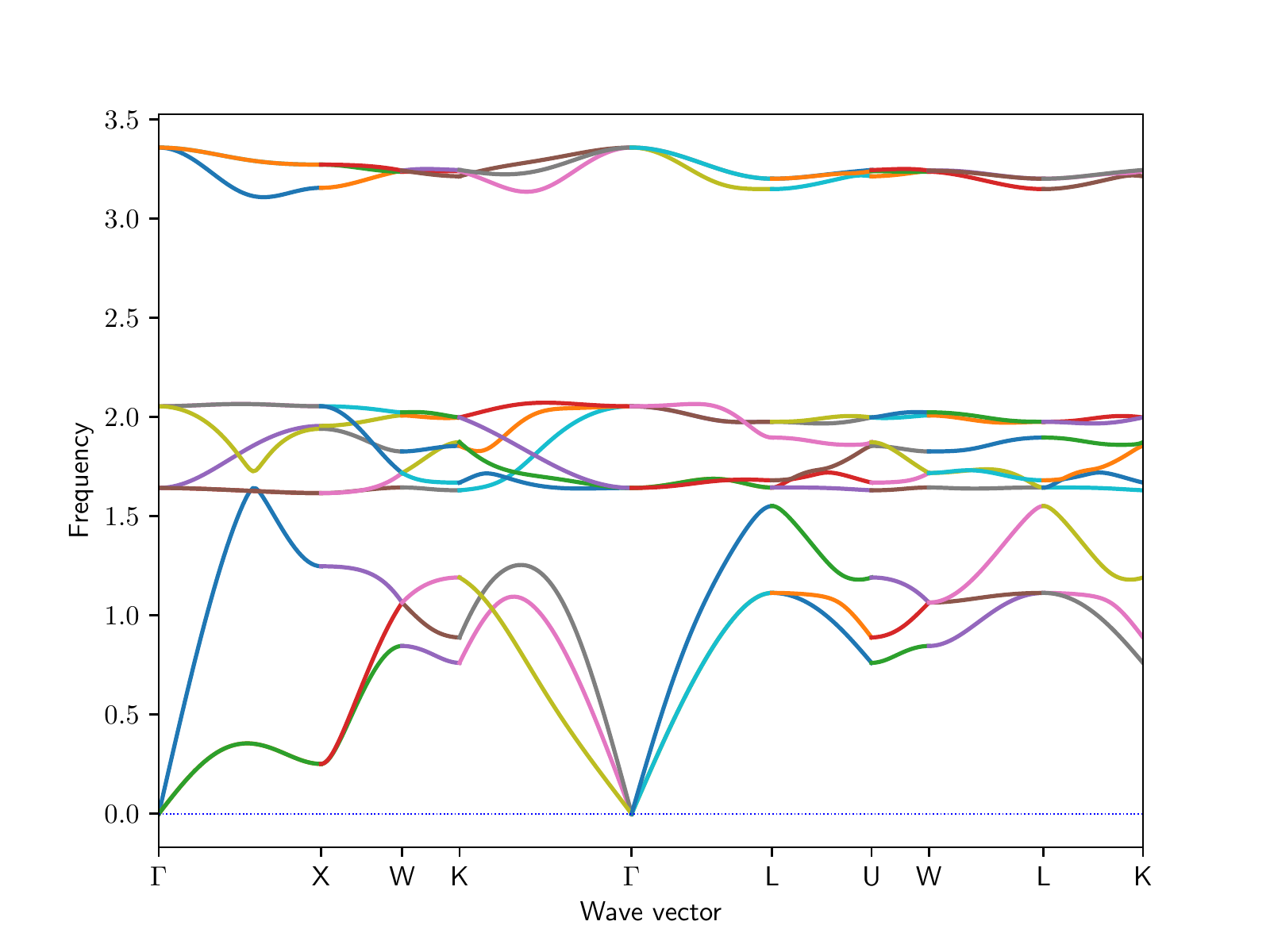}} \\
 \subfloat[Ag$_{2}$TmMg]{\includegraphics[width=0.55\columnwidth]{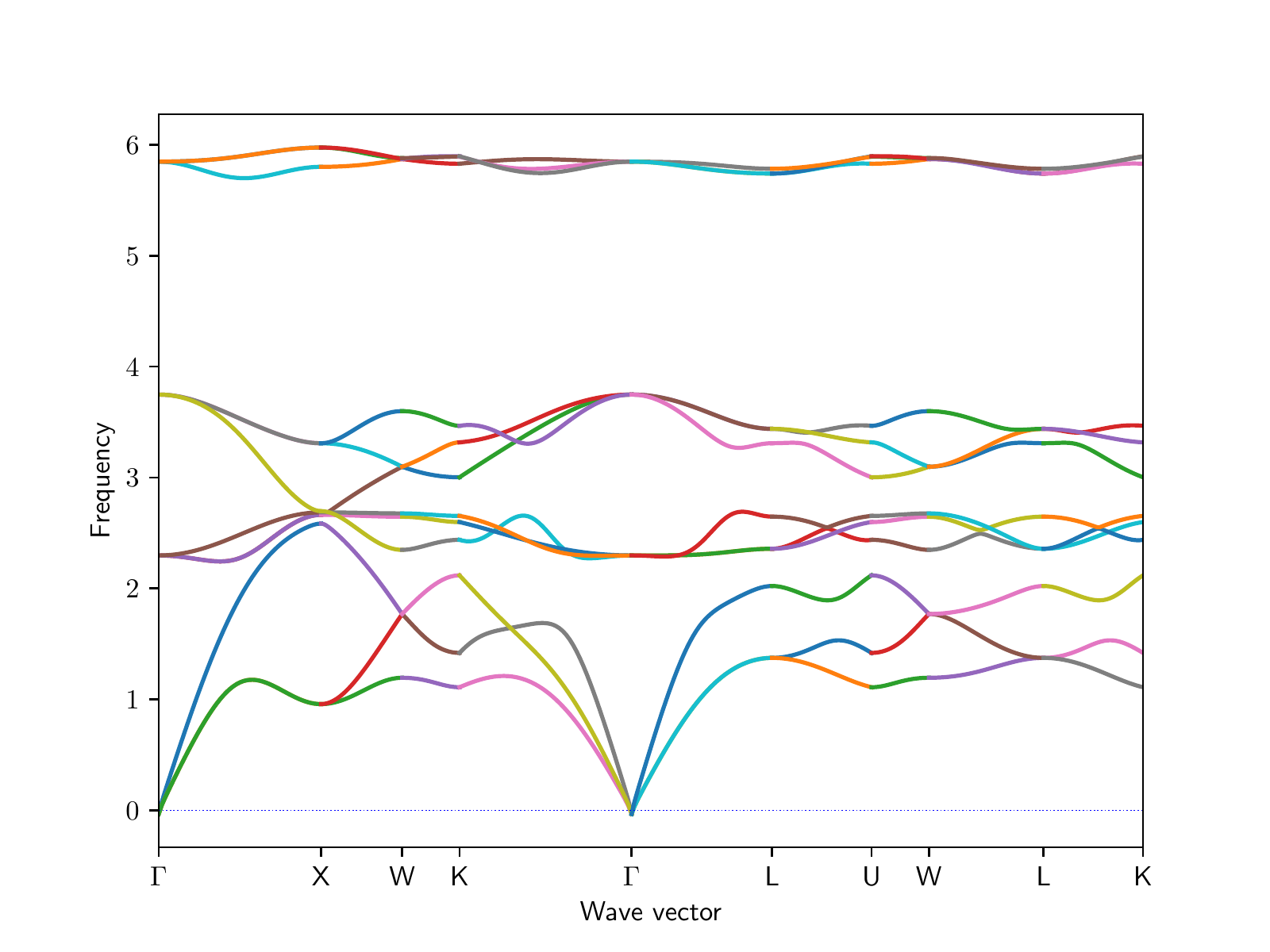}} 
 \caption{The phonon band structure of all full Heuslers compounds}
 \label{fig:phonon2}
\end{figure*}

\FloatBarrier
\section{Bandstructure}
  The electronic bandstructure of all the compounds studied without spin-orbit coupling is as shown:

\begin{figure*}[htb]
 \subfloat[Pd$_{2}$DyIn]{\includegraphics[width=0.55\columnwidth]{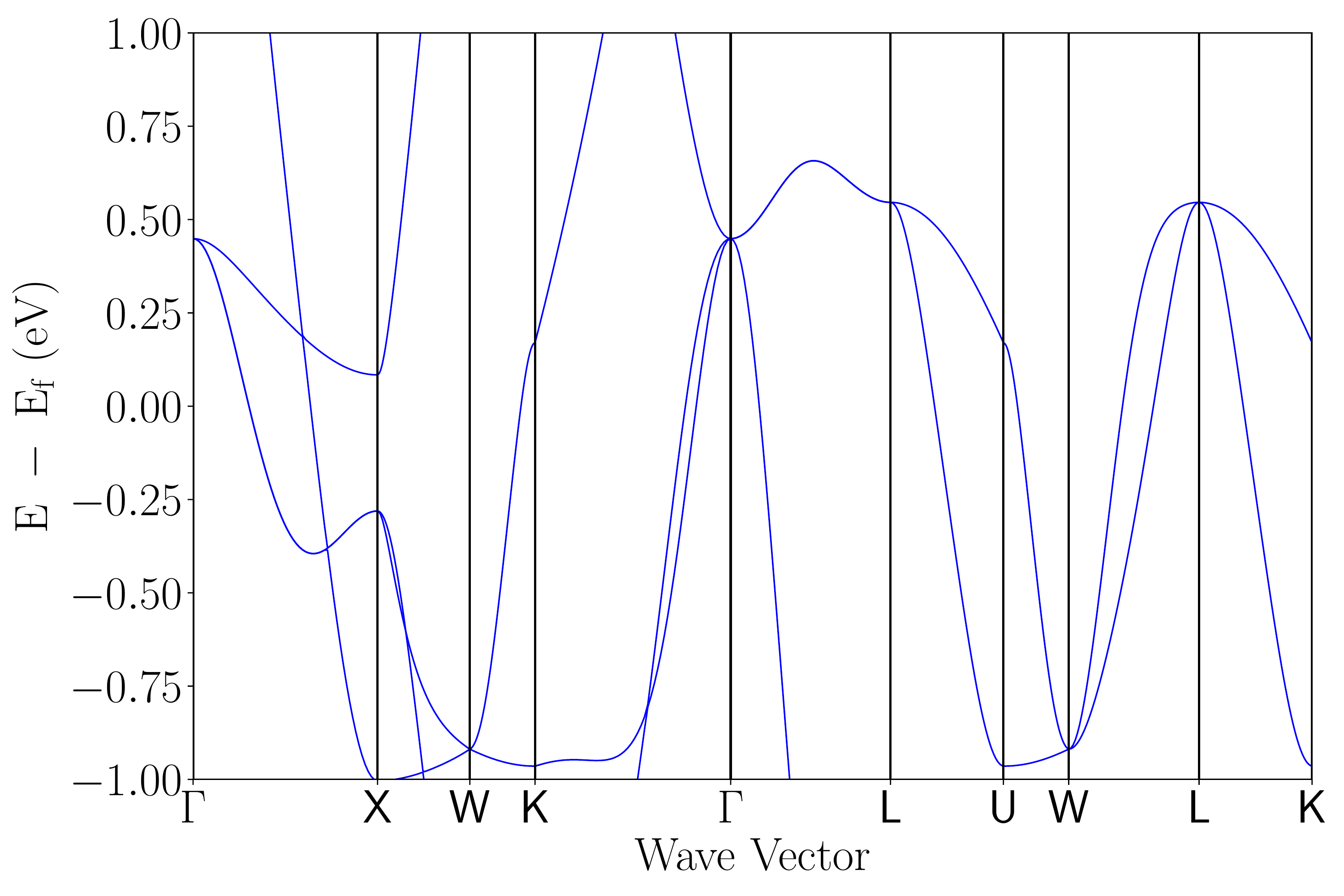}} 
 \subfloat[Hg$_{2}$LaMg]{\includegraphics[width=0.55\columnwidth]{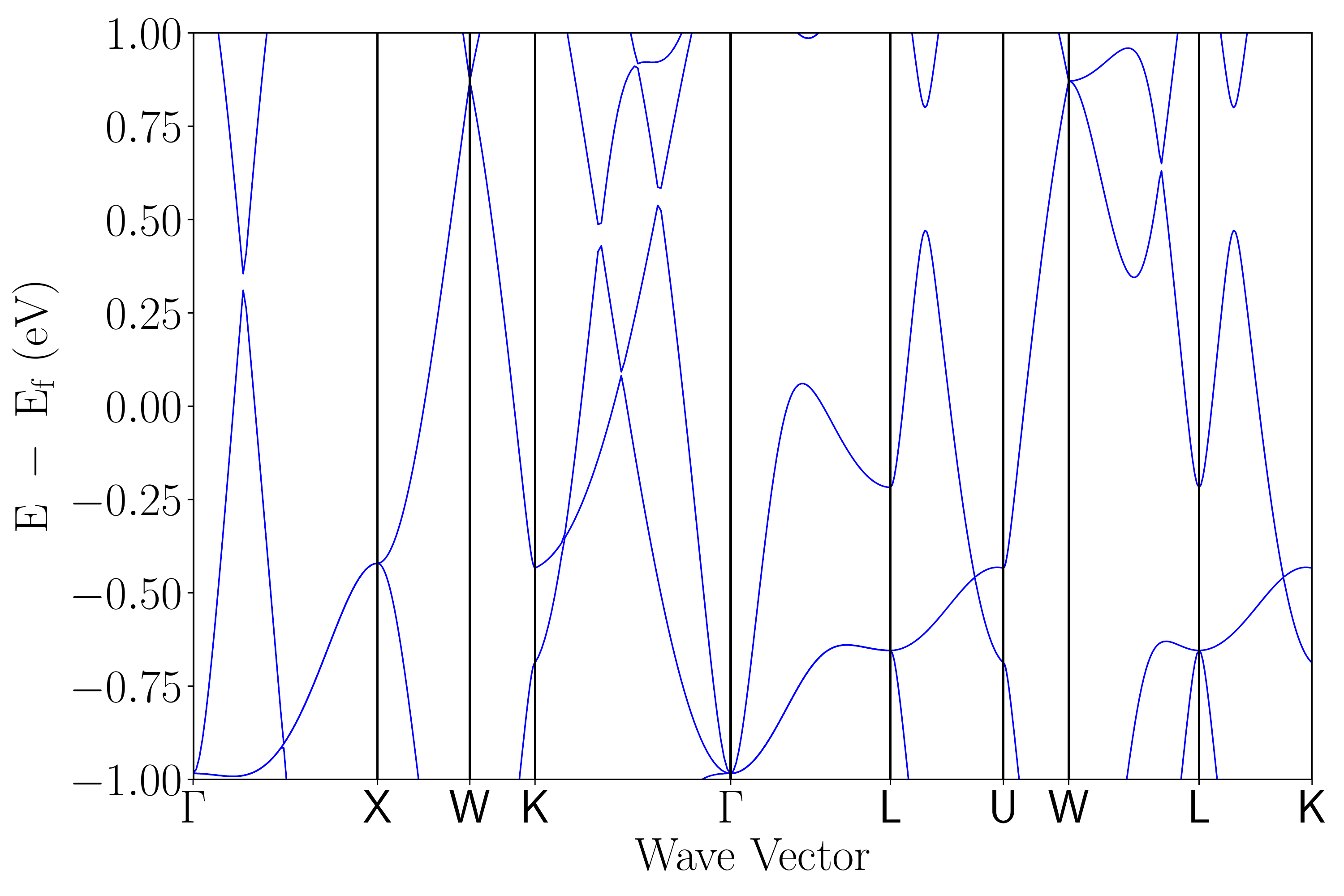}} \\
 \subfloat[Pd$_{2}$ErIn]{\includegraphics[width=0.55\columnwidth]{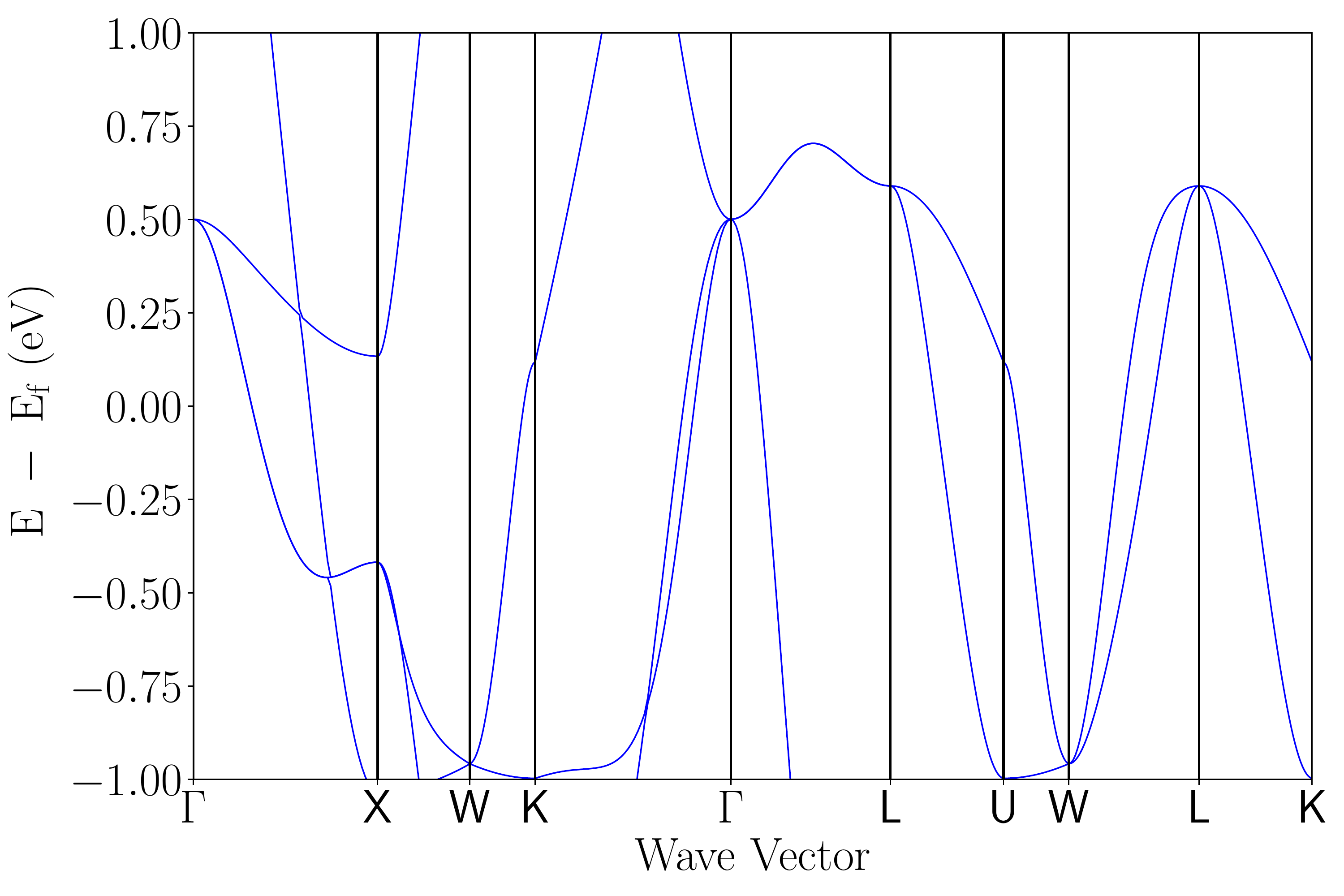}} 
 \subfloat[Hg$_{2}$PrAg]{\includegraphics[width=0.55\columnwidth]{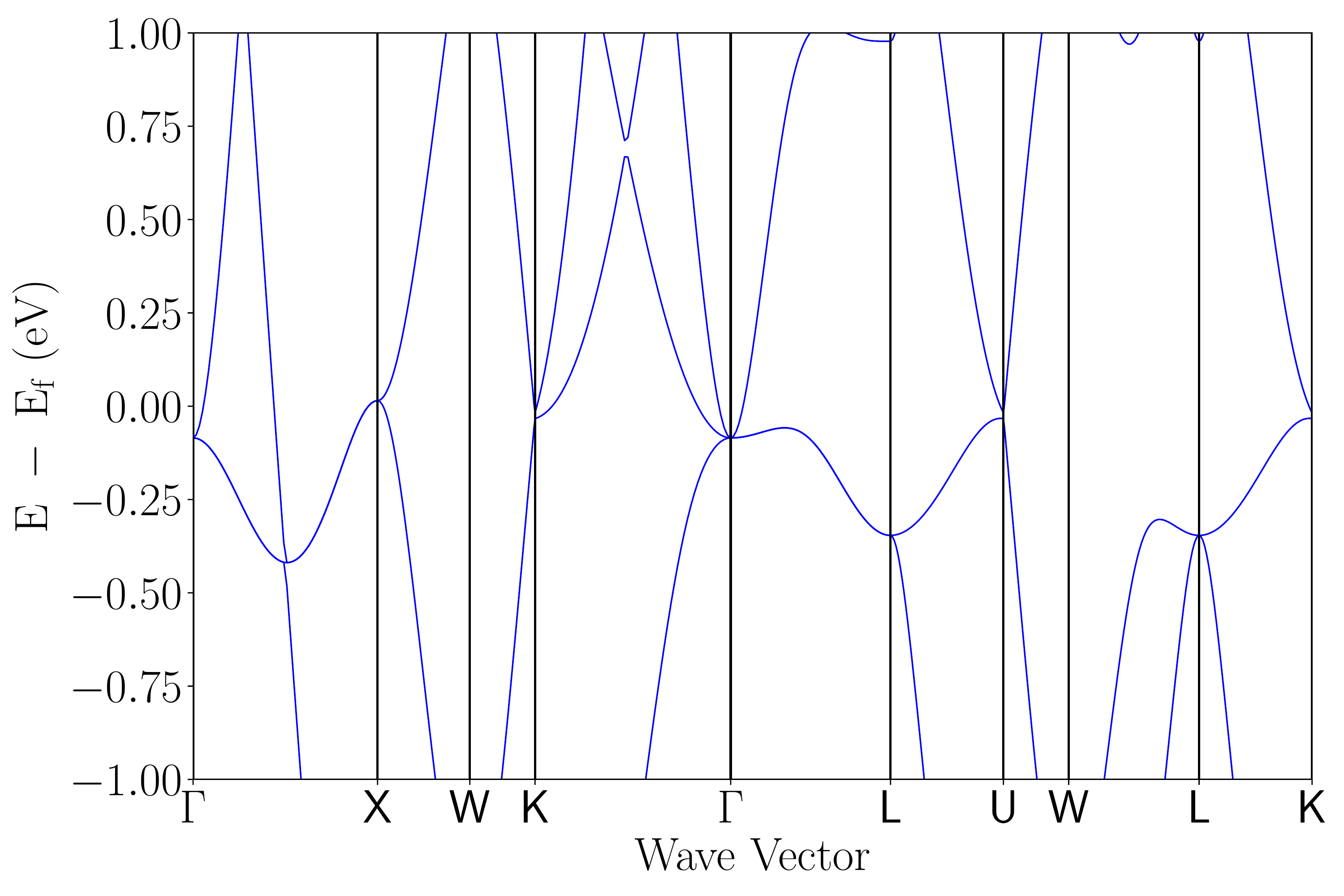}} \\
\end{figure*}

\begin{figure*}[htb]
 \ContinuedFloat
 \subfloat[Ag$_{2}$TbAl]{\includegraphics[width=0.55\columnwidth]{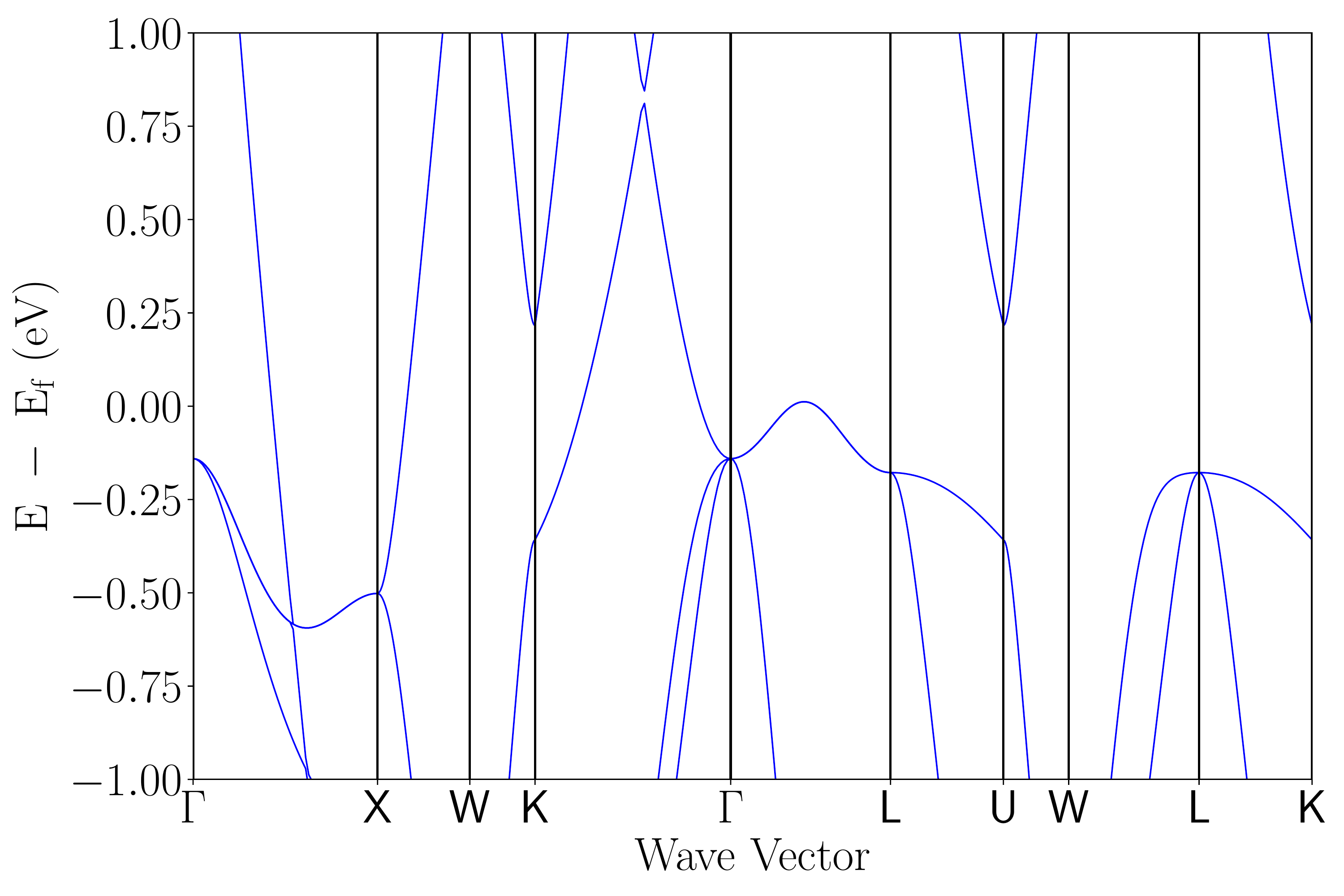}} 
 \subfloat[Ag$_{2}$YMg] {\includegraphics[width=0.55\columnwidth]{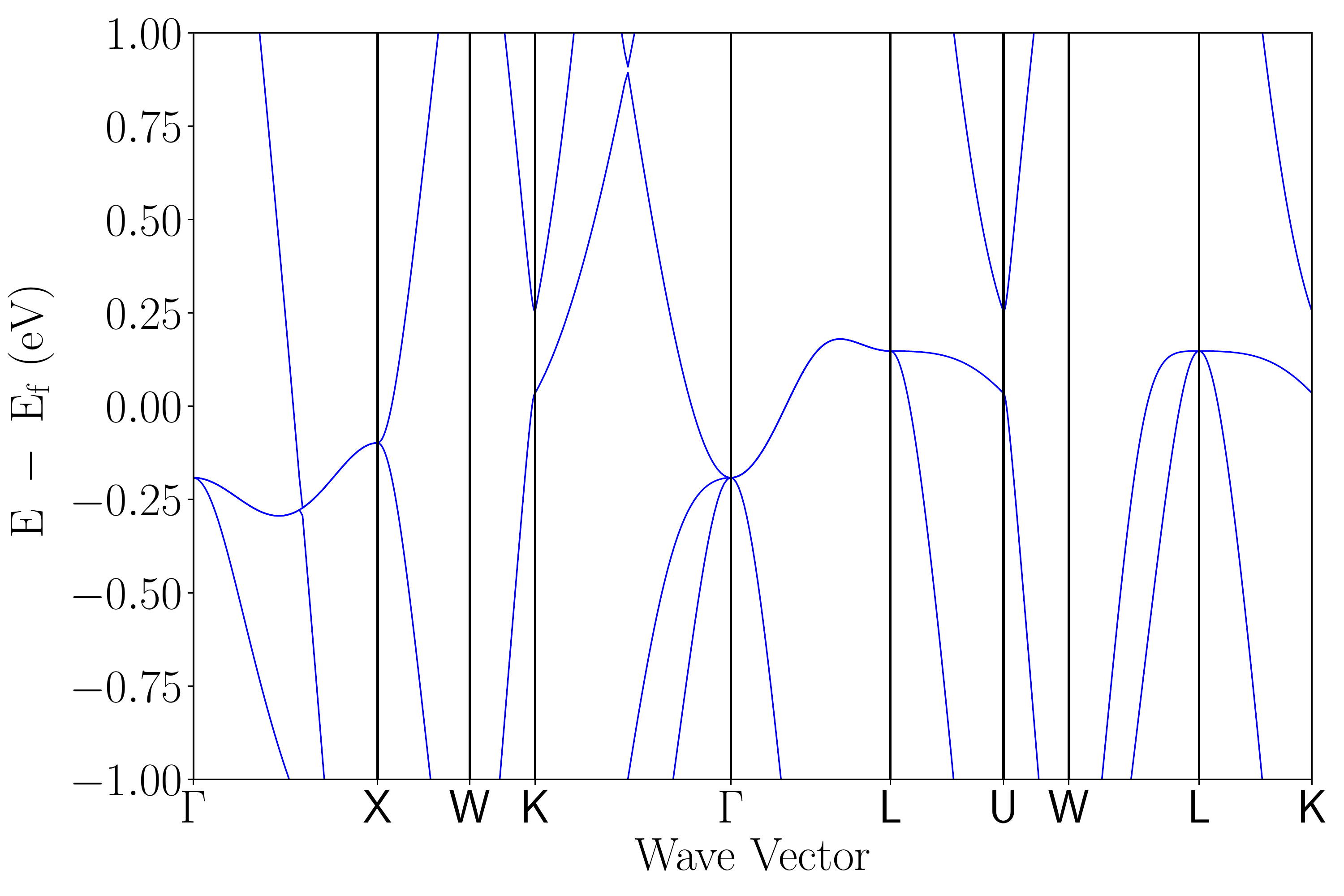}} \\
 \subfloat[Au$_{2}$DyMg]{\includegraphics[width=0.55\columnwidth]{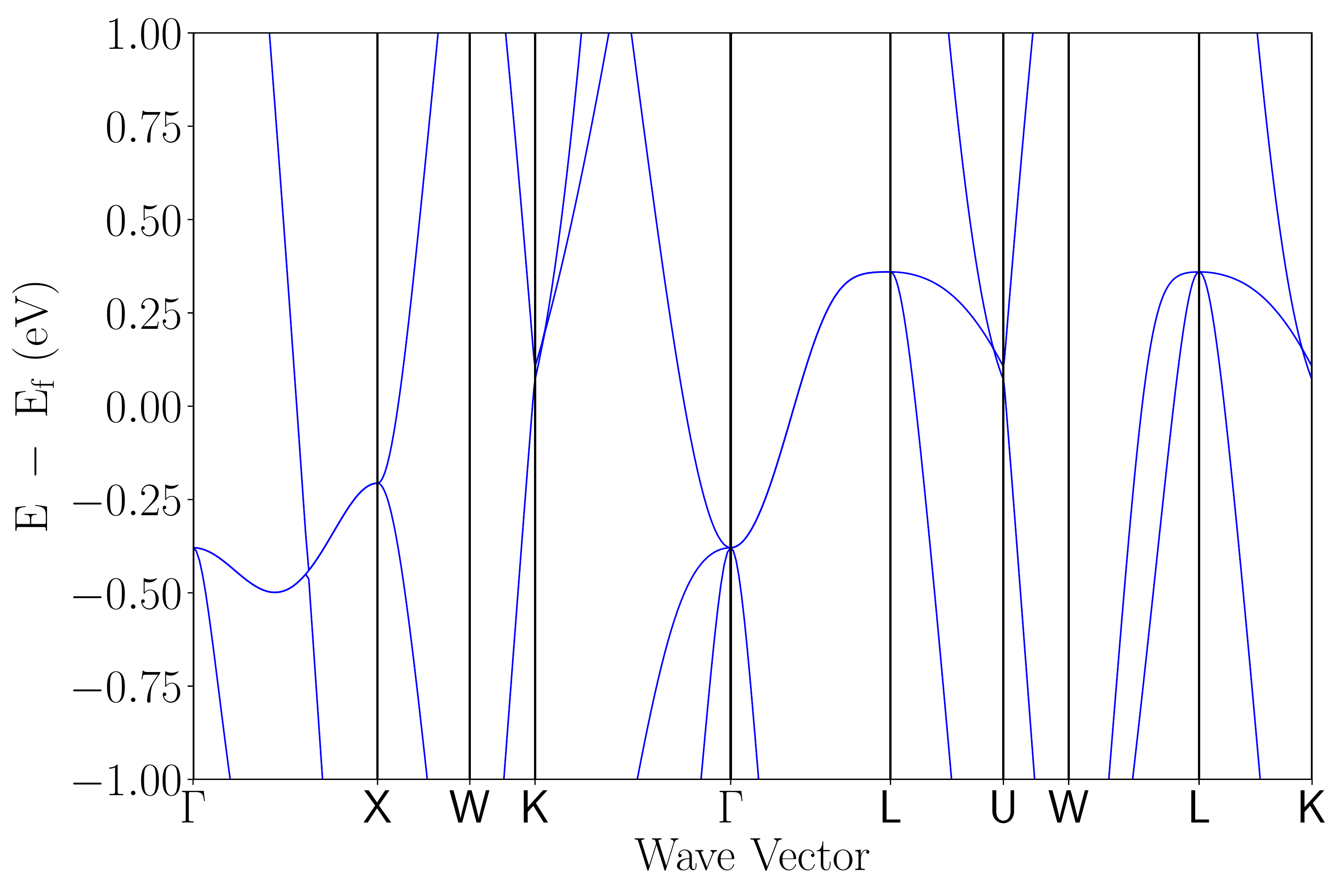}}
 \subfloat[Ag$_{2}$PrIn]{\includegraphics[width=0.55\columnwidth]{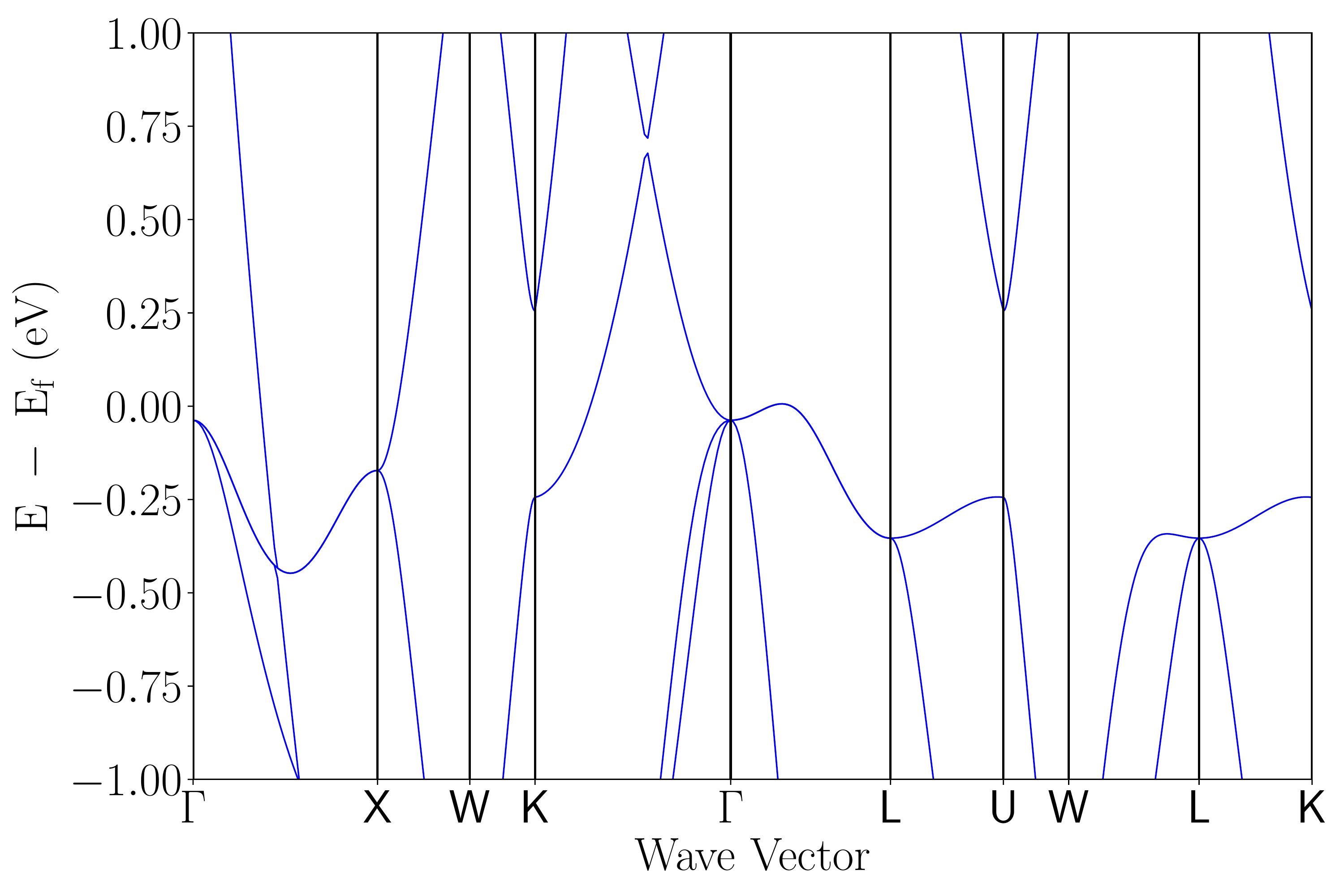}} \\
 \subfloat[Ag$_{2}$PrMg]{\includegraphics[width=0.55\columnwidth]{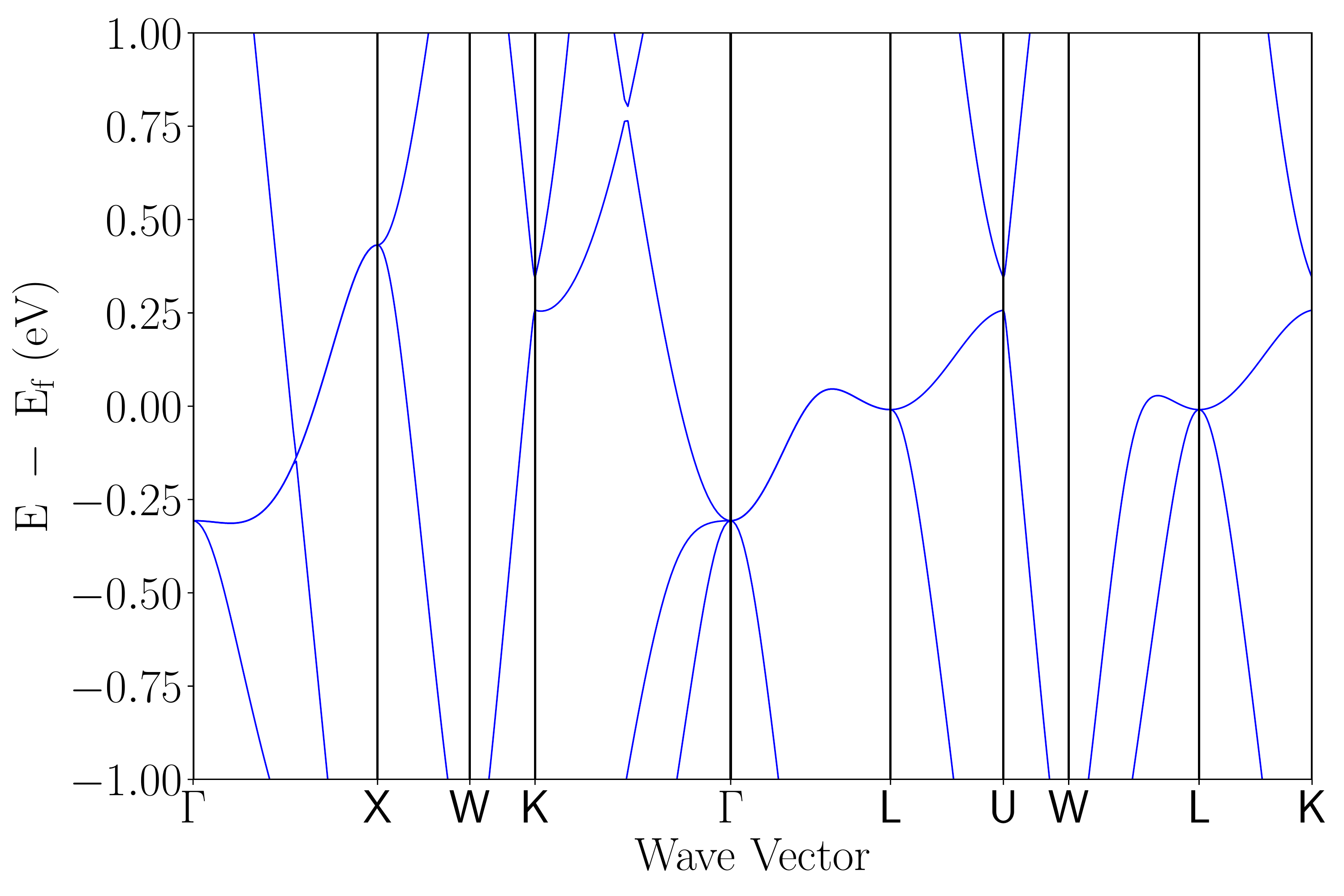}}
 \subfloat[Ag$_{2}$TbIn]{\includegraphics[width=0.55\columnwidth]{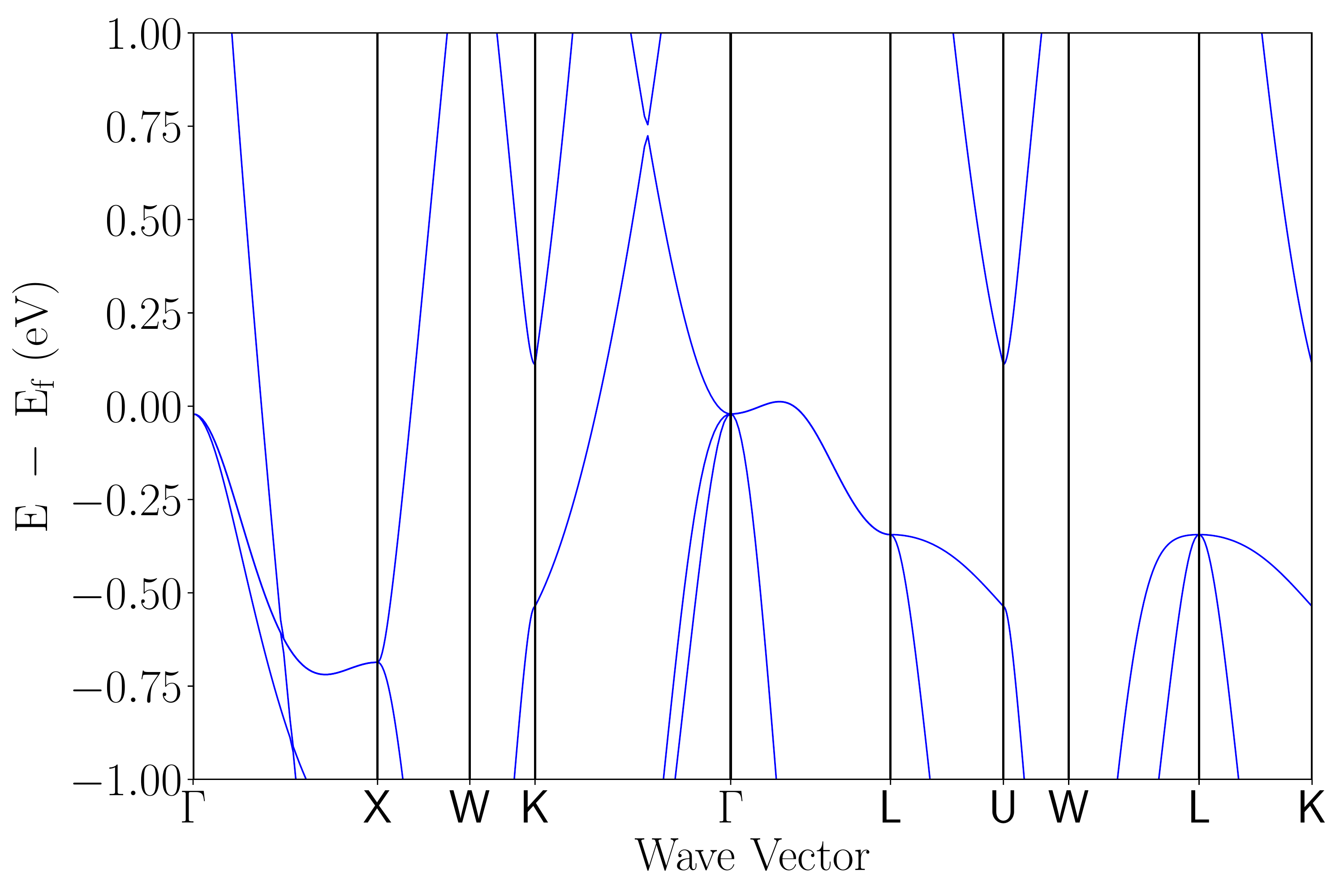}} \\
 \caption{The elelctronic band structure of all full Heuslers compounds}
 \label{fig:band1}
\end{figure*}

\begin{figure*}[htb]
 \subfloat[Au$_{2}$DyCd]{\includegraphics[width=0.55\columnwidth]{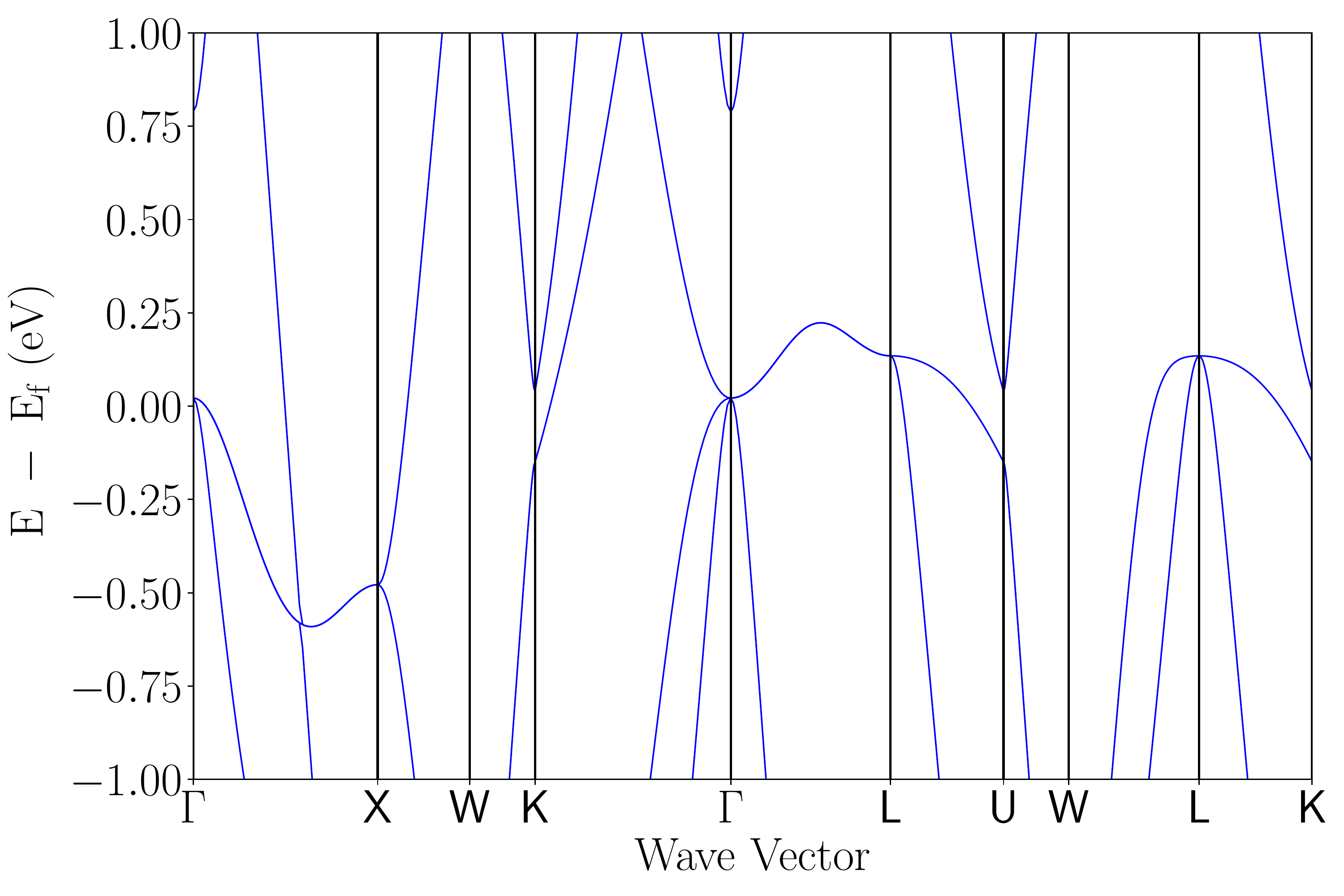}}
 \subfloat[Pd$_{2}$ScGa]{\includegraphics[width=0.55\columnwidth]{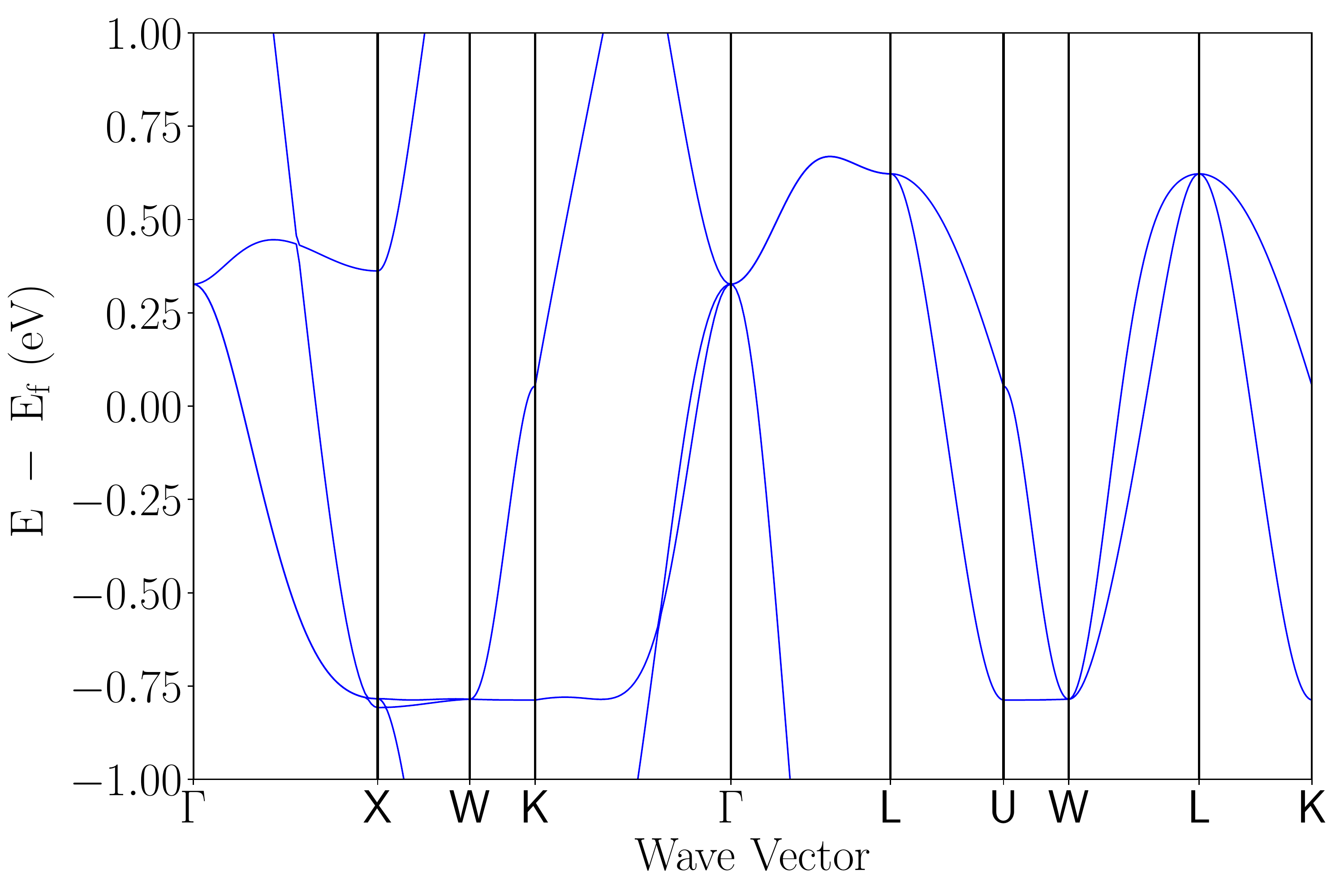}} \\ 
 \subfloat[Ag$_{2}$ScIn]{\includegraphics[width=0.55\columnwidth]{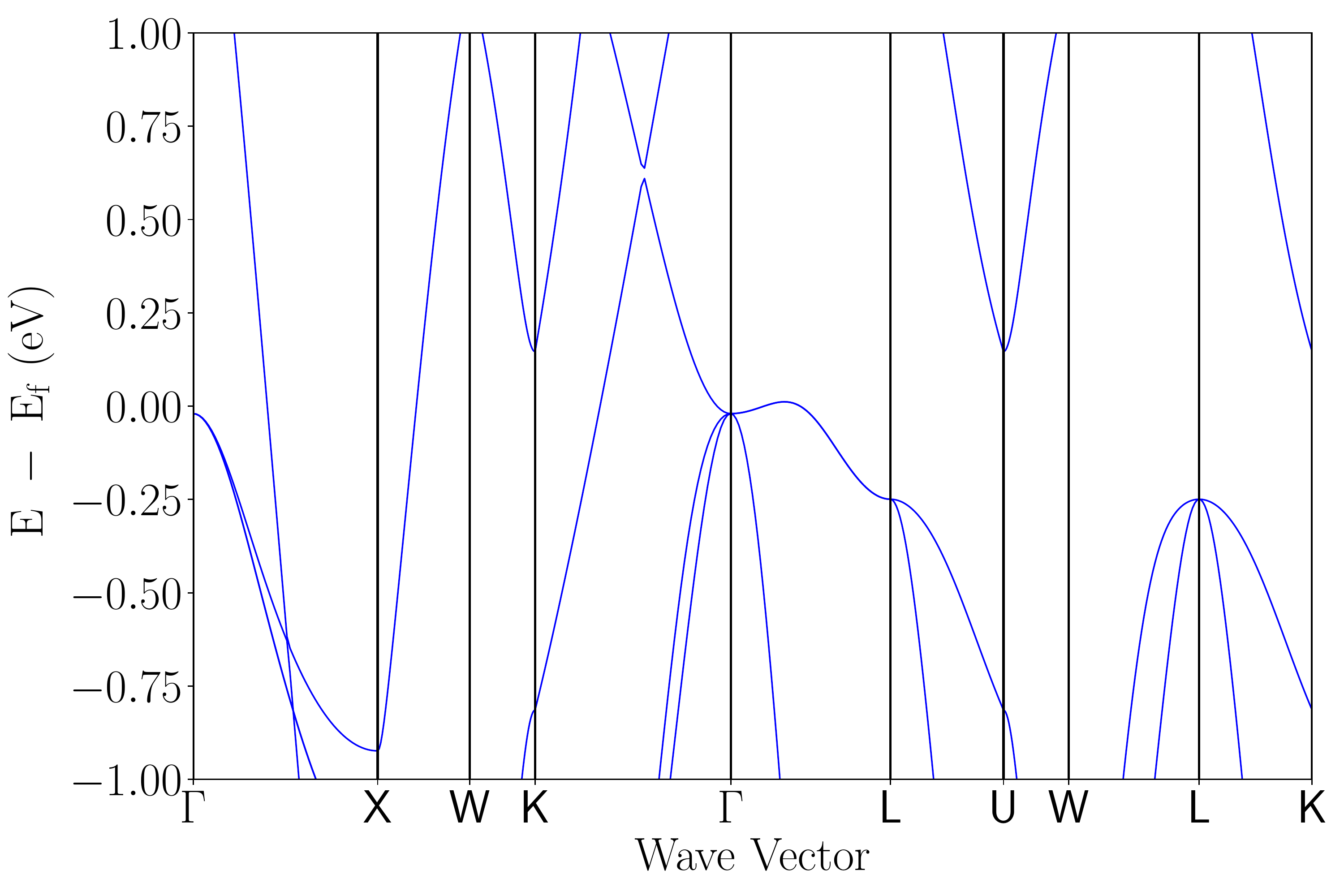}}
 \subfloat[Au$_{2}$PrIn]{\includegraphics[width=0.55\columnwidth]{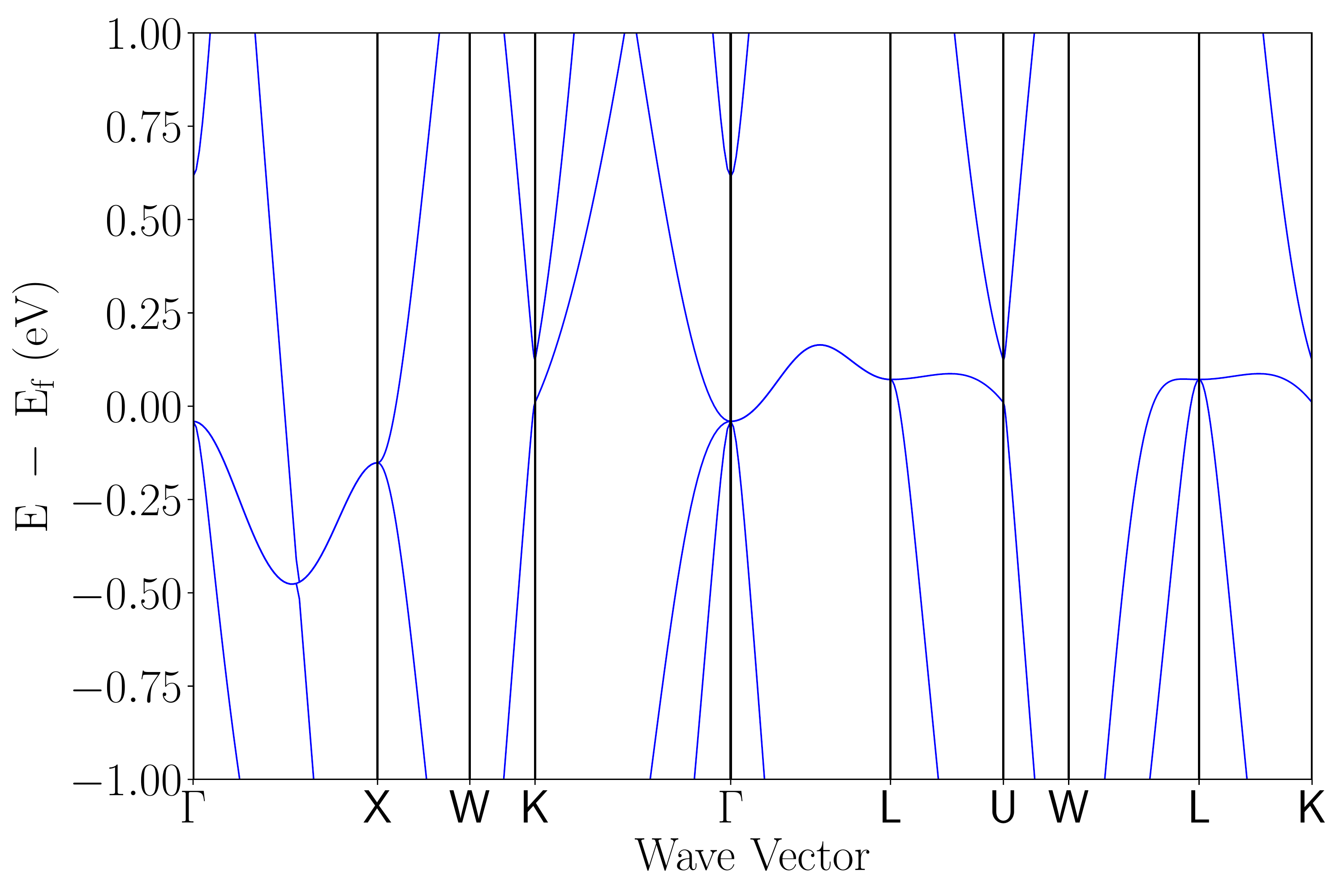}} \\
 \subfloat[Rh$_{2}$ZrAl]{\includegraphics[width=0.55\columnwidth]{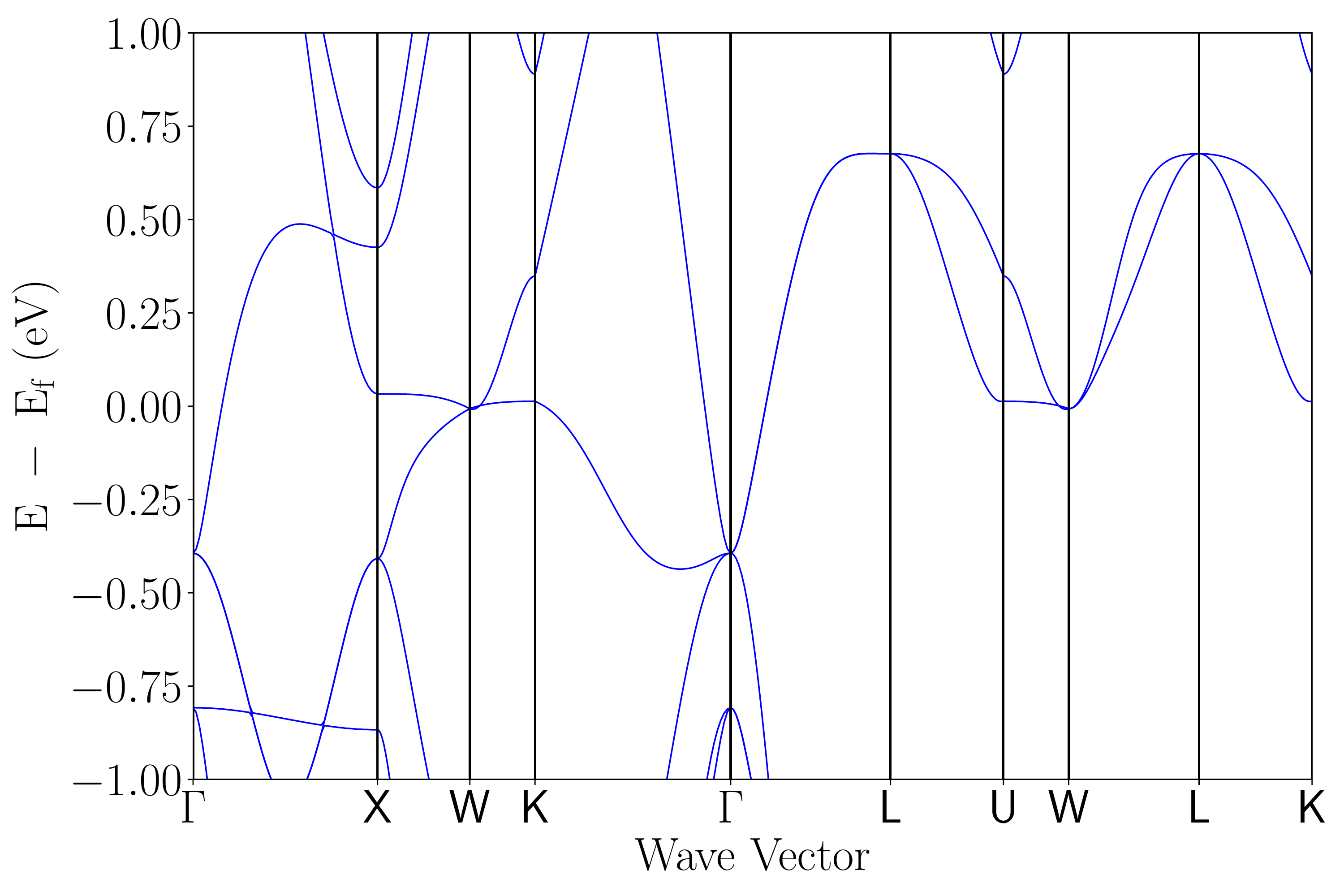}}
 \subfloat[Hg$_{2}$NaSm]{\includegraphics[width=0.55\columnwidth]{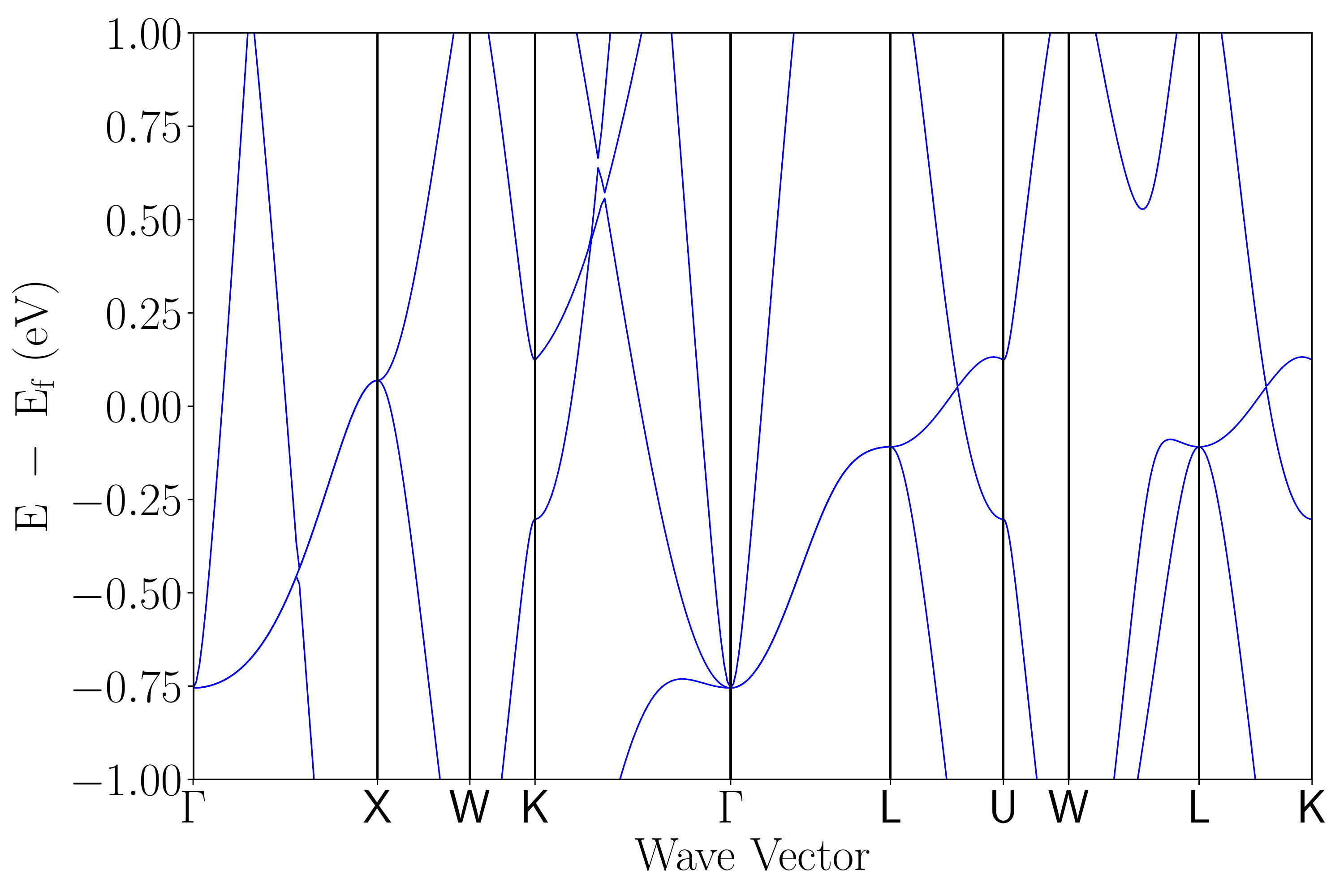}} \\ 
\end{figure*}

\begin{figure*}[htb]
 \ContinuedFloat
 \subfloat[Ag$_{2}$SmIn]{\includegraphics[width=0.55\columnwidth]{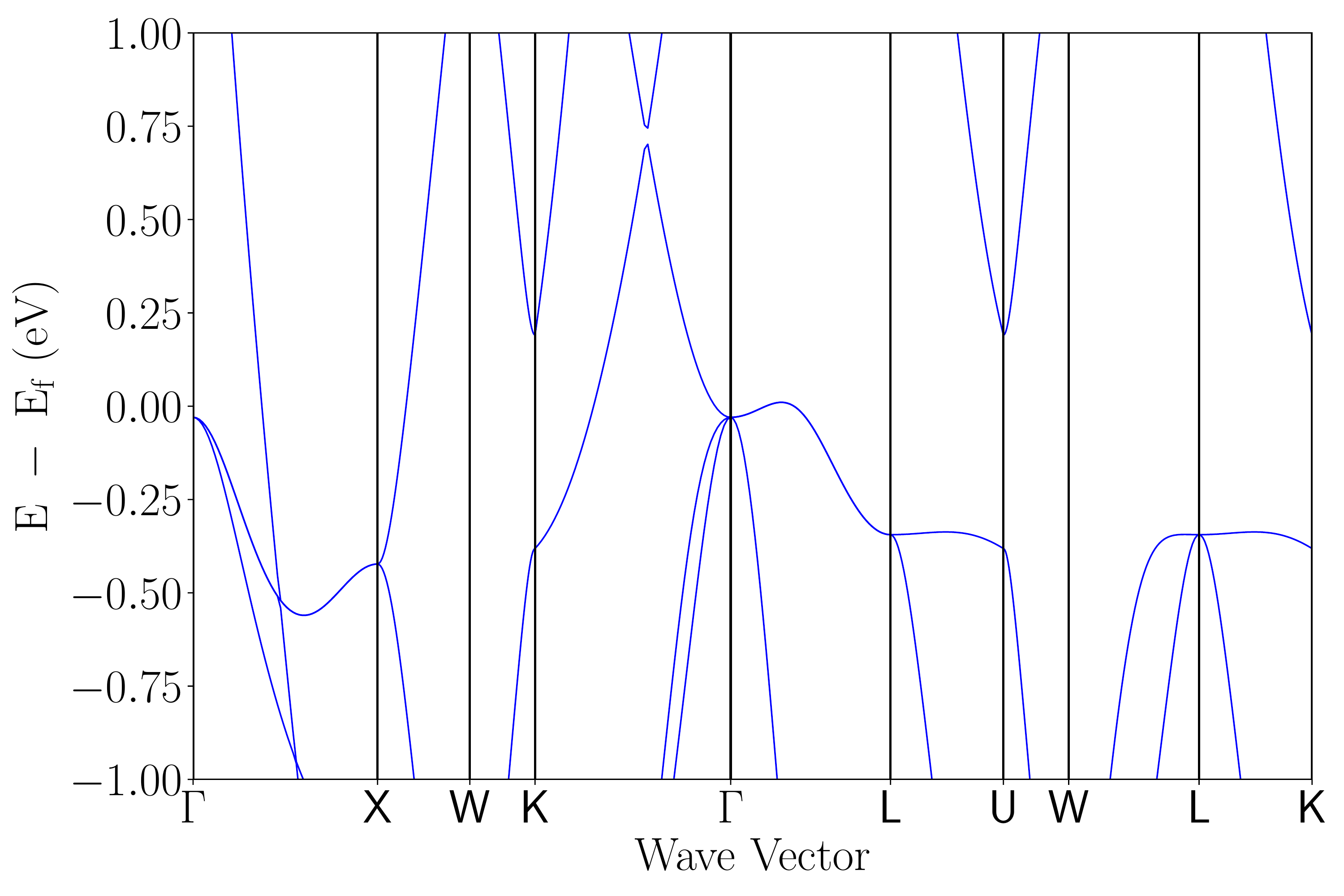}}
 \subfloat[Ag$_{2}$TmCa]{\includegraphics[width=0.55\columnwidth]{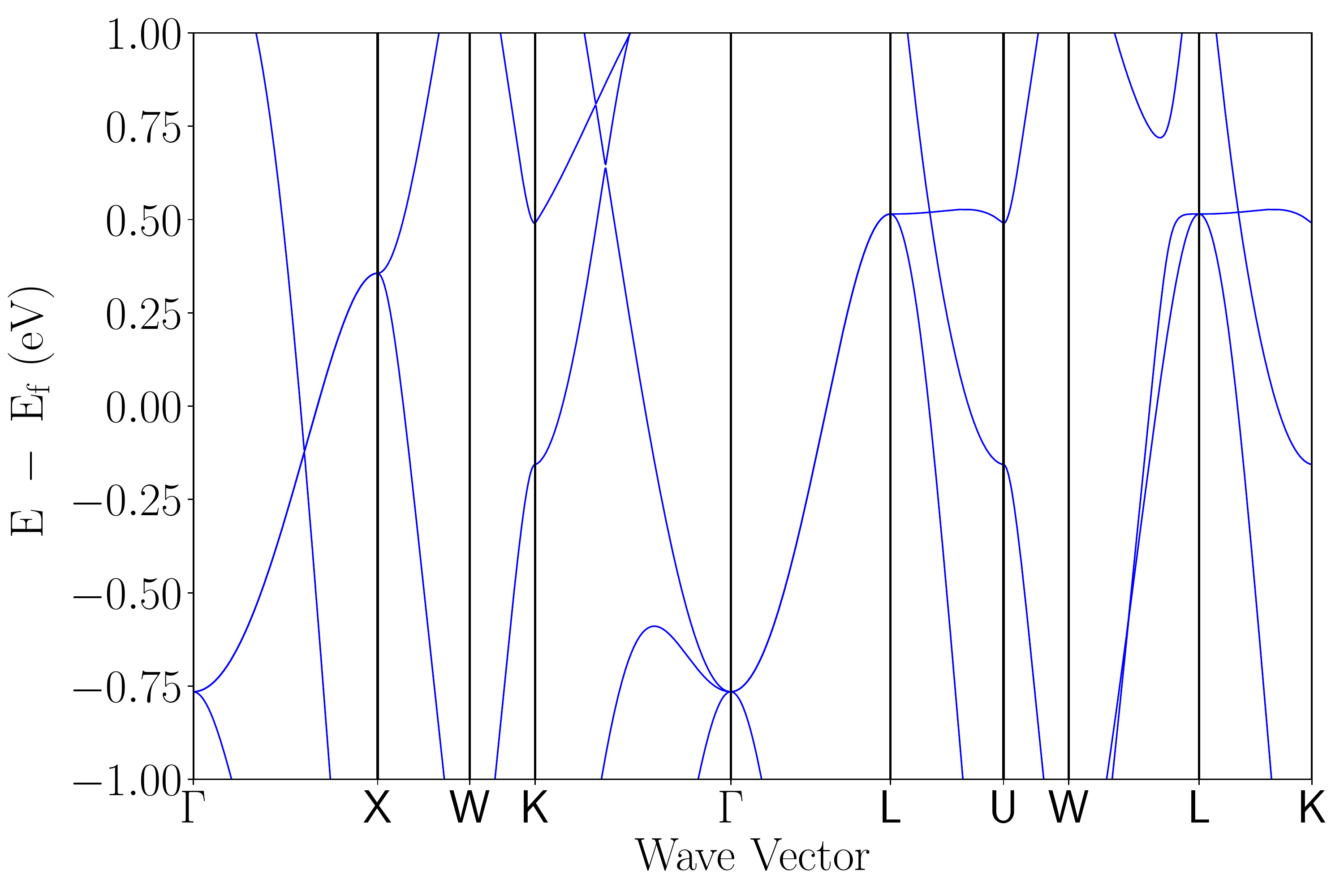}} \\
 \subfloat[Ag$_{2}$TmSr]{\includegraphics[width=0.55\columnwidth]{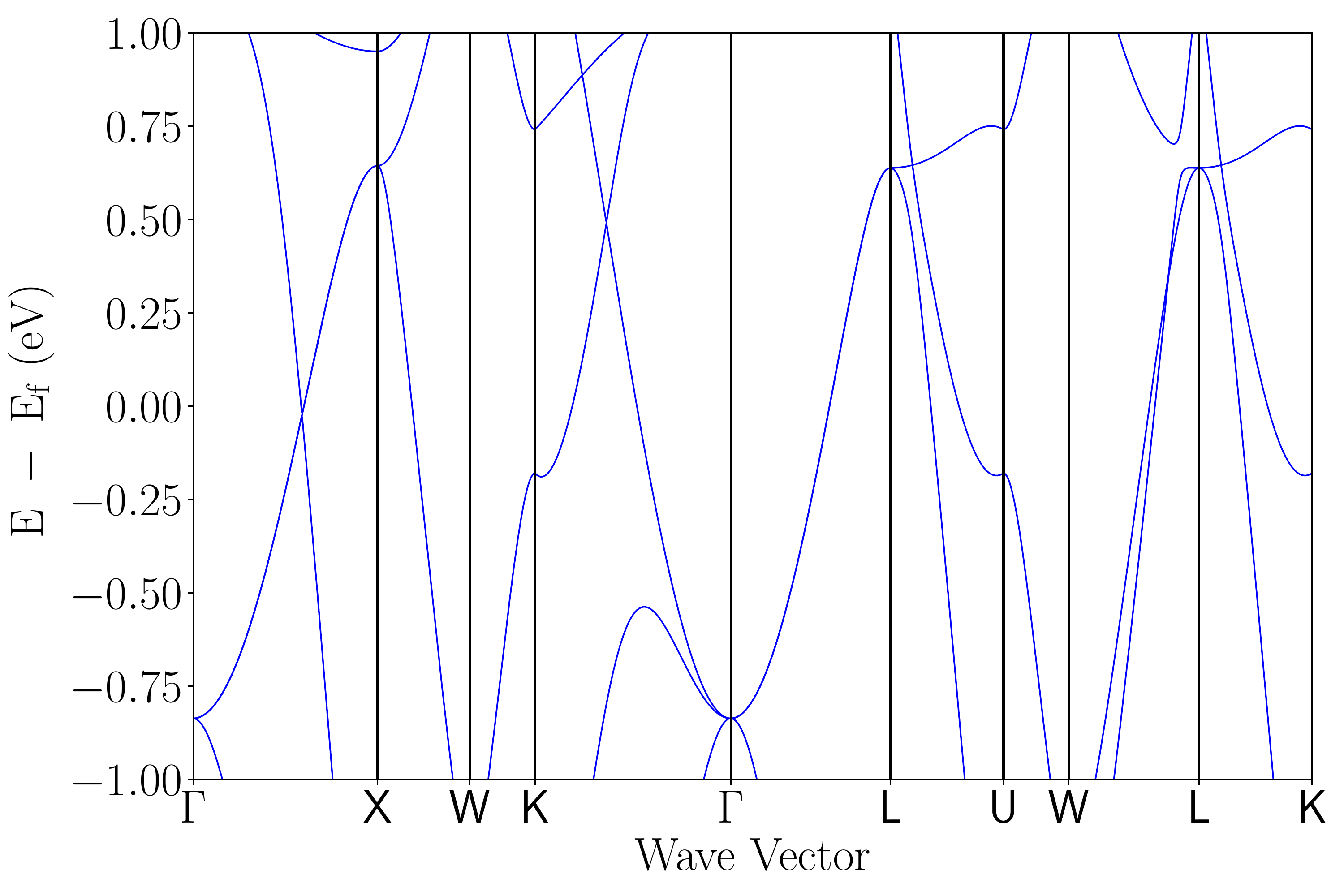}}
 \subfloat[Ag$_{2}$TmBa]{\includegraphics[width=0.55\columnwidth]{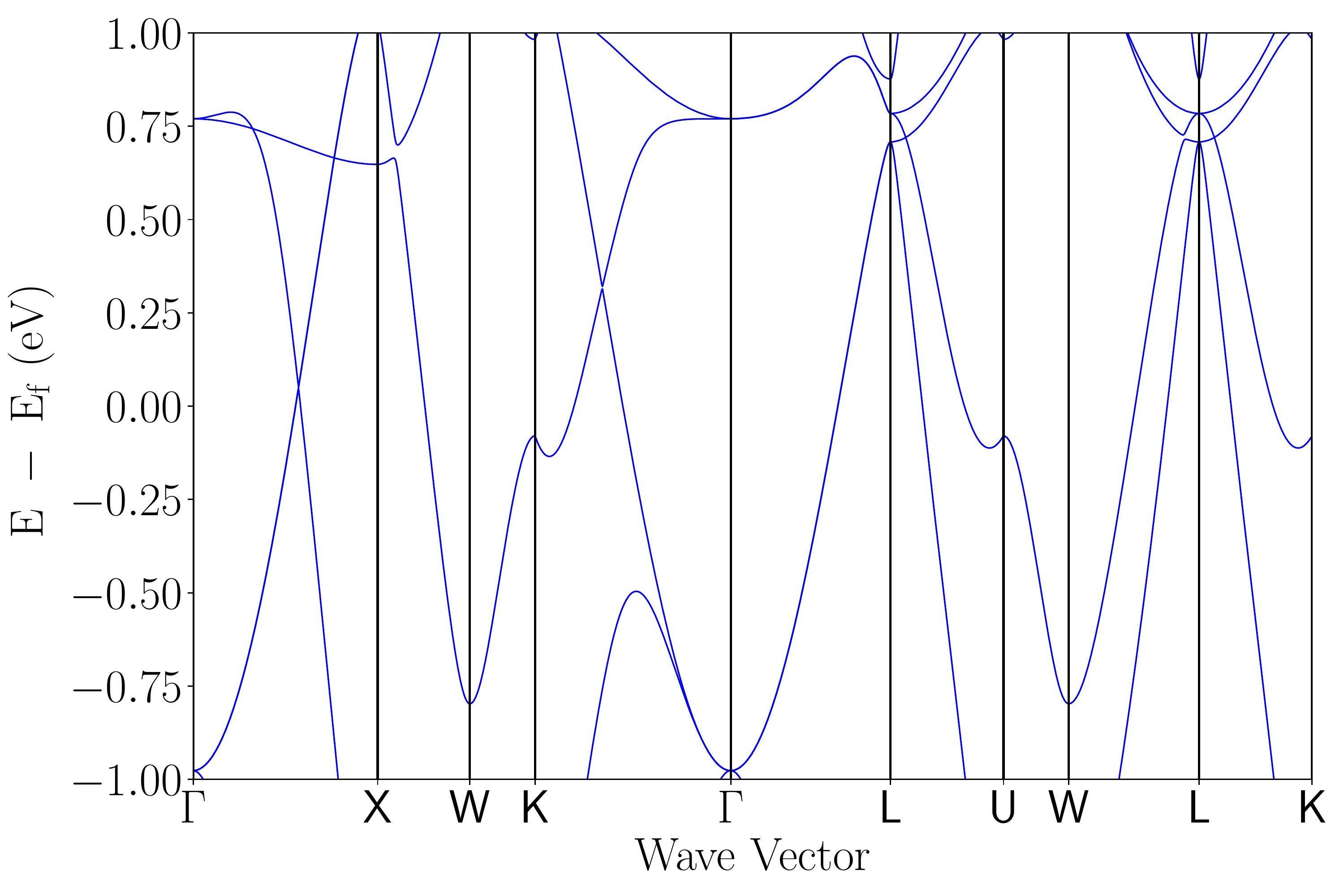}} \\
 \caption{The electronic band structure of all full Heuslers compounds}
 \label{fig:band2}
\end{figure*}